\documentclass[a4paper,usenatbib]{mnras}

\usepackage[T1]{fontenc}
\usepackage{ae,aecompl,appendix}
\usepackage[switch]{lineno}

\usepackage{graphicx}	
\usepackage{amsmath}	
\usepackage{amssymb}	
\usepackage{textcomp}
\usepackage{upgreek}
\usepackage{hyperref}
\usepackage{breakurl}
\usepackage{soul}

\newcommand{\urlwofont}[1]{\urlstyle{same}\url{#1}}

\title[SLSNe II from ZTF]{The Zwicky Transient Facility phase I sample of hydrogen-rich superluminous supernovae without strong narrow emission lines}

\author[T. Kangas et al.]{
\href{https://orcid.org/0000-0002-5477-0217}{T. Kangas},$^{1}$\thanks{E$-$mail: tuomask@kth.se} 
\href{https://orcid.org/0000-0003-1710-9339}{Lin Yan},$^{2}$  
\href{https://orcid.org/0000-0001-6797-1889}{S. Schulze},$^{3}$
\href{https://orcid.org/0000-0001-8532-3594}{C. Fransson},$^{4}$ 
\href{https://orcid.org/0000-0003-1546-6615}{J. Sollerman},$^{4}$
\href{https://orcid.org/0000-0001-9454-4639}{R. Lunnan},$^{4}$ 
\href{https://orcid.org/0000-0002-9646-8710}{C. M. B. Omand},$^{4}$ \newauthor
\href{https://orcid.org/0000-0002-8977-1498}{I. Andreoni},$^{5,6,7}$
R. Burruss,$^{8}$
\href{https://orcid.org/0000-0002-1066-6098}{T.-W. Chen},$^{4}$
A. J. Drake,$^{9}$
\href{https://orcid.org/0000-0002-4223-103X}{C. Fremling},$^{9}$
\href{https://orcid.org/0000-0002-3653-5598}{A. Gal-Yam},$^{10}$
\href{https://orcid.org/0000-0002-3168-0139}{M. J. Graham},$^{9}$ \newauthor
\href{https://orcid.org/0000-0001-5668-3507}{S. L. Groom},$^{11}$
J. Lezmy,$^{12}$
\href{https://orcid.org/0000-0003-2242-0244}{A. A. Mahabal},$^{9,13}$
\href{https://orcid.org/0000-0002-8532-9395}{F. J. Masci},$^{11}$ 
\href{https://orcid.org/0000-0001-8472-1996}{D. Perley},$^{14}$
\href{https://orcid.org/0000-0002-0387-370X}{R. Riddle},$^{2}$
\href{https://orcid.org/0000-0003-3433-1492}{L. Tartaglia},$^{4,15}$ \newauthor
and \href{https://orcid.org/0000-0001-6747-8509}{Y. Yao}$^{16}$
\\
$^{1}$Department of Physics, KTH Royal Institute of Technology, The Oskar Klein Centre, AlbaNova, SE-106 91 Stockholm, Sweden\\
$^{2}$The Caltech Optical Observatories, California Institute of Technology, Pasadena, CA 91125, USA \\
$^{3}$Department of Physics, The Oskar Klein Centre, Stockholm University, AlbaNova, SE-106 91 Stockholm, Sweden \\
$^{4}$Department of Astronomy, The Oskar Klein Centre, Stockholm University, AlbaNova, SE-106 91 Stockholm, Sweden \\
$^{5}$Joint Space-Science Institute, University of Maryland, College Park, MD 20742, USA \\
$^{6}$Department of Astronomy, University of Maryland, College Park, MD 20742, USA \\
$^{7}$Astrophysics Science Division, NASA Goddard Space Flight Center, Mail Code 661, Greenbelt, MD 20771, USA \\
$^{8}$Caltech Optical Observatories, California Institute of Technology, 35899 Canfield Road Palomar Mountain, CA 92060-0200 \\
$^{9}$Division of Physics, Mathematics, and Astronomy, California Institute of Technology, Pasadena, CA 91125, USA \\
$^{10}$Department of particle physics and astrophysics, Weizmann Institute of Science, 76100 Rehovot, Israel \\
$^{11}$IPAC, California Institute of Technology, 1200 E. California Blvd, Pasadena, CA 91225, USA \\
$^{12}$Universit\'{e} de Lyon, Universit\'{e} Claude Bernard Lyon 1, CNRS/IN2P3, IP2I Lyon, F-69622, Villeurbanne, France \\
$^{13}$Center for Data Driven Discovery, California Institute of Technology, Pasadena, CA 91125, USA \\
$^{14}$Astrophysics Research Institute, Liverpool John Moores University, 146 Brownlow Hill, Liverpool L3 5RF \\
$^{15}$INAF - Osservatorio Astronomico di Padova, Vicolo dell’Osservatorio 5, I-35122 Padova, Italy \\
$^{16}$Cahill Center for Astronomy and Astrophysics, California Institute of Technology, Pasadena, CA 91125, USA \\
}
\date{Accepted XXX. Received YYY; in original form ZZZ}
\pubyear{2021}

\begin{document}
\label{firstpage}
\pagerange{\pageref{firstpage}$-$\pageref{lastpage}}
\maketitle

\begin{abstract}

We present a sample of 14 hydrogen-rich superluminous supernovae (SLSNe~II) from the Zwicky Transient Facility (ZTF) between 2018 and 2020. We include all classified SLSNe with peaks $M_{g}<-20$~mag and with observed \emph{broad} but not narrow Balmer emission, corresponding to roughly 20 per cent of all hydrogen-rich SLSNe in ZTF phase I. We examine the light curves and spectra of SLSNe~II and attempt to constrain their power source using light-curve models. The brightest events are photometrically and spectroscopically similar to the prototypical SN 2008es, while others are found spectroscopically more reminiscent of non-superluminous SNe~II, especially SNe~II-L. $^{56}$Ni decay as the primary power source is ruled out. Light-curve models generally cannot distinguish between circumstellar interaction (CSI) and a magnetar central engine, but an excess of ultraviolet (UV) emission signifying CSI is seen in most of the SNe with UV data, at a wide range of photometric properties. Simultaneously, the broad H$\alpha$ profiles of the brightest SLSNe~II can be explained through electron scattering in a symmetric circumstellar medium (CSM). In other SLSNe~II without narrow lines, the CSM may be confined and wholly overrun by the ejecta. CSI, possibly involving mass lost in recent eruptions, is implied to be the dominant power source in most SLSNe~II, and the diversity in properties is likely the result of different mass loss histories. Based on their radiated energy, an additional power source may be required for the brightest SLSNe~II, however -- possibly a central engine combined with CSI.

\end{abstract}

\begin{keywords}
transients: supernovae $-$ stars: magnetars $-$ stars: mass-loss $-$ galaxies: statistics
\end{keywords}



\section{Introduction}

Massive stars ($\gtrsim8~\mathrm{M}_\odot$) end their lives in supernova (SN) explosions. In addition to ordinary core-collapse SNe (CCSNe), wide-field, untargeted transient searches have also uncovered classes of superluminous SNe \citep[SLSNe; for reviews see][]{galyam12,galyam19}; these are analogous to hydrogen-poor (i.e. Type I) and hydrogen-rich (Type II) CCSNe, but reach peak absolute magnitudes $\lesssim-20$ mag \citep[e.g.][]{quimby11,decia18,chen22a}, even possibly $\sim-23$ mag \citep[][]{dong16,li20} -- but see also \citet{leloudas16}. The source of this unusual luminosity is as yet debated. Mechanisms commonly considered for SLSNe include the decay of $^{56}$Ni synthesized in the SN explosion, requiring amounts of $^{56}$Ni unattainable except in pair-instability SNe (PISNe) of extremely massive stars \citep{barkat67,hw02}; interaction with circumstellar matter (CSM) efficiently converting the kinetic energy of the ejecta into radiation \citep[e.g.][]{cf94,ofek07,sorokina16}; and a central engine such as a strongly-magnetized, fast-spinning neutron star born in the collapse -- a millisecond magnetar -- that provides additional energy to the ejecta \citep[][]{kb10,woosley10}. Another possible central engine is fallback accretion onto a nascent black hole \citep{dexterkasen13}. Depending on the mechanism, SLSNe are generally thought to require very massive progenitors (whether single or binary), initially on the order of $\gtrsim30~\mathrm{M}_\odot$ or even $\gtrsim100~\mathrm{M}_\odot$ \citep[e.g.][]{jerkstrand17,lunnan18,stevance21}.

Type I SLSNe have been studied more extensively in the literature than Type II, owing to more abundant observational data \citep[see e.g.][]{moriya18,quimby18,chen22a}. A magnetar central engine is commonly invoked as the power source \citep[e.g.][]{nicholl13,nicholl15,inserra13}, although circumstellar interaction (CSI) models are a better match to some SLSNe~I \citep[][]{lunnan18,chen22b}. Undulations in SLSN light curves \citep[][]{chen17,inserra17} are often best explained by CSI {\citep[see e.g.][]{west22}} despite the lack of strong narrow emission lines, but variations in the central engine remain a plausible mechanism for some objects as well \citep[e.g.][]{chugaiutrobin21,chen22b, moriya22}.

Type II SLSNe, the prototype of which is SN 2006gy \citep{ofek07,smith07}, mostly exhibit classic Type IIn spectra, and can be considered simply the bright end of the Type IIn luminosity function \citep[e.g.][]{perley16}. These ``SLSNe~IIn" are most likely powered by CSI. There are, however, SLSNe II without strong narrow lines as well. The prototype of this hitherto small group of events is SN 2008es \citep{gezari09,miller09}, which bore a strong similarity to {linearly declining SNe II, historically called SNe II-L \citep{barbon79}.} Typical SNe~II-L have recently been considered analogous to the fainter, slower SNe~II-P \citep{valenti15}, but with more massive progenitors \citep[possibly 15--20~$\mathrm{M}_\odot$;][]{vandyk99,faran14,kangas17a}, higher mass loss rates, less hydrogen in the ejecta and {therefore a short or nonexistent plateau phase \citep{anderson14,bose18,reynolds20}.}

A subset of SNe II-L, such as SN 1979C and SN 2013fc, are unusually luminous, reaching peak absolute magnitudes of $\sim-20$~mag \citep{panagia80,devauc81,kangas16}. {SN 1998S \citep[e.g.][]{leonard00,fassia01} and a few other SNe~IIn are similar to these and could be considered SNe~II-L with early or weak narrow lines \citep{taddia13,inserra13b,tartaglia21}.} While the light curves of most SNe II seem to require some early CSI \citep{morozova18}, the properties of the luminous SNe II-L suggest stronger CSI is responsible for boosting their luminosity above that of normal SNe II \citep[e.g.][]{kangas16}. CSI has also been suggested to power the prototypical broad-lined SLSN~II, SN 2008es, which resembles a SN~II-L, but with a 15--20 d delay in its spectroscopic evolution, somewhat weaker absorption lines and a broader light curve \citep{gezari09,miller09}. 

\citet{inserra18} studied SN 2008es and two other SLSNe~II that lacked strong narrow emission lines. Until now, these three SNe and CSS121015 \citep{benetti14} have been the only such SNe known in the literature. Following \citet{inserra18} and \citet{galyam19}, for the purposes of this paper we will use the term 'SLSNe~II' to refer to only such SNe. SLSNe with strong narrow Balmer lines will be called SLSNe IIn. A magnetar central engine, analogous to those possibly powering SLSNe I, was found compatible with SLSN~II light curves and temperatures by \citet{inserra18}. On the other hand, CSI can produce a variety of emission line profiles, not necessarily narrow \citep[e.g.][]{moriya12,mcdowell18,taddia20}. Late-time observations of SN 2008es favoured the CSI scenario, but a magnetar central engine has not been excluded as the \emph{dominant} power source of SLSNe II \citep{kornpob19}. The host galaxies of the known members of this subclass are similar to those of SLSNe I \citep{inserra18}, i.e. blue, faint and metal-poor \citep[while SLSNe IIn show a wider range in host properties;][]{perley16,Schulze2021a}.

In recent years, the Zwicky Transient Facility \citep[ZTF;][]{bellm19,graham19} has detected a multitude of transients due to its moderately high-cadence and untargeted mapping of the entire northern sky down to 20.5 mag. The SNe observed by the ZTF collaboration in its first phase (March 2018 -- October 2020) include sufficient numbers of SLSNe II to examine this subclass in greater quantity than previously done and to shed more light on the source of its often-extreme luminosity and ambiguous power source. To that end, in this paper we present a sample of 14 SLSNe II from ZTF's phase I with broad Balmer lines but without strong narrow emission lines, examine their properties and fit their light curves using semi-analytical models of $^{56}$Ni decay, magnetar spin-down and CSI.

We describe the sample and the ZTF survey itself in Sect.~\ref{sec:data}. We examine the spectroscopic evolution of the sample events in Sect.~\ref{sec:spec} and their light curves in Sect.~\ref{sec:lc}, while comparing them to the objects in \citet{inserra18}. We describe the light-curve modelling process and its results in Sect.~\ref{sec:models}, and examine the host galaxies of our sample in Sect.~\ref{sec:hosts}. We discuss our findings and possible power sources in Sect.~\ref{sec:disco}, and present our conclusions in Sect.~\ref{sec:concl}. Throughout this paper, magnitudes are in the AB system \citep{okegunn83}, and $\Lambda$CDM cosmological parameters are assumed to be $H_0 = 69.6$~km~s$^{-1}$~Mpc$^{-1}$, $\Omega_M = 0.286$ and $\Omega_\Lambda = 0.714$ \citep{bennett14}.

\section{Sample and data reduction}
\label{sec:data}

\begin{table*}
\begin{minipage}{0.8\linewidth}
\centering
    \caption{Properties of the SLSNe in our sample.}
    \begin{tabular}{lcccccc}
       ZTF name & IAU name & RA & Decl. & $z$ & $A_{V,\mathrm{Gal}}$ & Classification \\
        & & (J2000, h:m:s) & (J2000, deg:m:s) & &  (mag) & \\
        \hline
        ZTF18acsxwdi & SN~2018jkq & 01:09:31.53 & +29:19:52.7 & 0.119 & 0.166 & SLSN II\\
        ZTF19aalvdeu & SN~2019kwr & 13:22:12.35 & +49:54:52.7 & 0.202 & 0.035 & SLSN II\\
        ZTF19aamrais & SN~2019cqc & 18:21:43.05 & +30:59:33.5 & 0.117 & 0.323 & SLSN II\\
        ZTF19aavakzo & SN~2019gsp & 22:10:04.24 & +23:28:38.5 & 0.171 & 0.231 & SLSN II\\
        ZTF19abxequc & SN~2019xfs & 19:10:04.90 & +32:20:41.4 & 0.116 & 0.513 & SLSN II\\
        ZTF19abxgmzr & SN~2019pud & 21:12:55.00 & -16:38:07.1 & 0.114 & 0.199 & SLSN I.5\\
        ZTF19acblhej & SN~2018lqi & 02:18:21.94 & -25:54:24.5 & 0.202 & 0.032 & SLSN II\\
        ZTF19ackiwff & SN~2019aanx & 07:29:59.99 & +13:05:43.2 & 0.403 & 0.235 & SLSN II\\
        ZTF19ackzvdp & SN~2019uba & 02:23:28.74 & -01:58:59.0 & 0.304 & 0.074 & SLSN II\\
        ZTF19adcfsoc & SN~2019zcr & 12:58:42.92 & +15:12:42.1 & 0.260 & 0.063 & SLSN I.5\\
        ZTF20aajvyja & SN~2020bfe & 17:57:50.69 & +33:47:48.1 & 0.099 & 0.120 & SLSN II\\
        ZTF20aatqene & SN~2020hgr & 14:16:26.51 & +70:24:48.8 & 0.126 & 0.042 & SLSN II\\
        ZTF20aayprqz & SN~2020jhm & 15:33:02.28 & +67:54:48.4 & 0.057 & 0.088 & SLSN I.5\\
        ZTF20acnznms & SN~2020yue & 11:00:00.32 & +21:06:45.8 & 0.204 & 0.054 & SLSN II \\
        \hline
    \end{tabular}
    \label{tab:objects}
\end{minipage}
\end{table*}  

In the ZTF Northern Sky Public Survey \citep{bellm2019b} of ZTF, all fields with center declination $\rm \delta \ge -31^\circ$ and galactic latitude $\rm \mid b \mid\, > 7^\circ$ (i.e. $\sim23,675$\,deg$^2$) are covered every three nights. This is roughly the entire Northern sky accessible from Palomar. Our target selection process is briefly described below. For more details, see \citet{chen22a}. 

A filter algorithm selects promising SLSN candidates from among ZTF alerts, which are then visually examined by human scanners. The filter excludes moving targets, stars, Galactic-plane targets and bogus alerts. The filter also prefers faint, blue host galaxies and long rise times in order to minimize non-SLSN contamination. Nearly all candidates ($\geq$95 per cent) brighter than 18.5 mag are classified by the ZTF Bright Transient Survey \citep[BTS;][]{fremling2020}. The rest, if not already classified on the Transient Name Server\footnote{\url{https://www.wis-tns.org/}}, were classified by the ZTF SLSN team and followed up. 

Instruments used to observe the spectra in this paper were the Spectral Energy Distribution Machine \citep[SEDM;][]{Blagorodnova2018} and the Double Beam Spectrograph \citep[DBSP;][]{okegunn83} on the Palomar 60\,inch (P60) and 200\,inch (P200) telescope respectively; the Low Resolution Imaging Spectrometer \citep[LRIS;][]{Oke1995} on the Keck I telescope; the Alhambra Faint Object Spectrograph and Camera (ALFOSC) on the 2.56m Nordic Optical Telescope (NOT); the Intermediate-dispersion Spectrograph and Imaging System (ISIS) on the 4.2m William Herschel Telescope (WHT); and the SPectrograph for the Rapid Acquisition of Transients (SPRAT) on the 2m robotic Liverpool Telescope \citep[LT;][]{Steele04}. The log of spectroscopic observations used here is available as supplementary material. A sample of the log is presented in Table \ref{apptab:Log}. 

The ZTF phase I ran from March 17, 2018 to {November 30}, 2020. During this period, a total of 63 SNe were discovered by the survey and classified as SLSNe~II (note that this classification includes SLSNe~IIn). The ZTF phase I also includes six events classified ``SLSNe~I.5"; i.e. brighter than $-20$ mag and showing early spectroscopic similarity to SLSNe I, or simply {lacking hydrogen lines until later} during the photospheric phase. {We define our sample as follows:}
\begin{itemize}
    \item Classified on the GROWTH Marshal \citep{kasliwal19} as either SLSN I.5 or SLSN II, i.e. including hydrogen Balmer lines and brighter than $-20$ mag {before $K$-correction (a total of 69 SNe)};
    \item Exhibiting a broad H$\alpha$ emission line -- full width half maximum (FHWM) of the broad component $\geq$5000~km~s$^{-1}$ -- in at least one spectrum {(20 SNe out of 69)};
    \item Lacking a typical Type IIn H$\alpha$ profile {(i.e. a multi-component profile with a narrow component from the SN)} past the light curve peak {(14 SNe out of 20)}. Narrow features {that disappear after the peak} were ignored, as CSI lines are seen in many very early spectra of normal SNe II \citep{khazov16,bruch21}. 
\end{itemize}

\begin{figure}
\centering
\includegraphics[width=0.95\columnwidth]{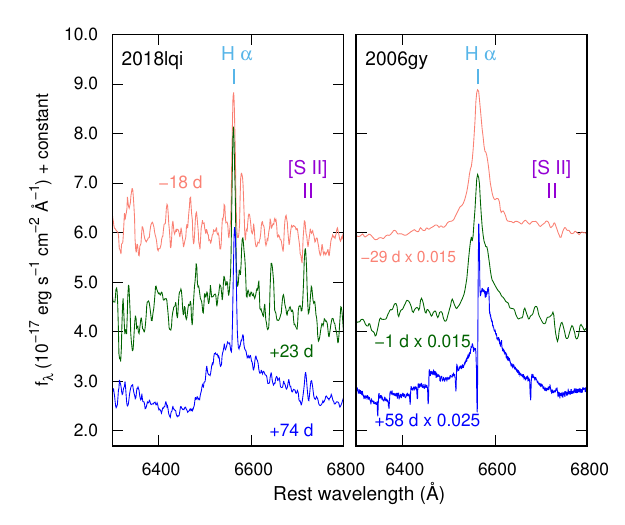}
\caption{H$\alpha$ profile of SN~2018lqi at different epochs (left), compared to the profile in the superluminous Type~IIn SN~2006gy \citep[right;][]{smith10}. {A broad component is clearly visible, but the unresolved narrow component is attributable to the host galaxy.} A Type IIn profile, on the other hand, consists of a narrow core with Lorentzian wings.}
\label{fig:haexample}
\end{figure}

These criteria are based on the spectra from ALFOSC, DBSP or LRIS, {as very low-resolution spectra} cannot reliably be used to detect narrow emission lines. Most objects in the sample show weak narrow emission {lines that seem to originate in the host galaxy.} Changing line ratios between H$\alpha$ and [O~{\sc iii}]~$\lambda\lambda4959,5007$ and [S~{\sc ii}]~$\lambda\lambda6717,6730$, or clearly Lorentzian profiles of the narrow Balmer line(s), imply an origin in the SN itself; otherwise an origin in the host galaxy is possible and we include the SN in our sample. The profiles of the narrow components are relatively weak, unresolved and (when applicable) similar to the [O~{\sc iii}] or [S~{\sc ii}] profiles. As an example of a target with such behaviour, we show the H$\alpha$ profile of SN~2018lqi and its evolution in Fig. \ref{fig:haexample}. We note that for SN~2018jkq, we only have a late-time (+170~d) spectrum with the broad H$\alpha$ emission, but both its early and late spectra strongly resemble other SNe in this sample. 

{The final sample of 14 SLSNe and their basic properties are listed in Table \ref{tab:objects}. The redshifts of the objects range from $0.057$ to $0.403$.} The quality of the follow-up varies widely, in terms of both spectroscopic cadence and the number of photometric filters used. {As such, the narrow lines in two SLSNe IIn in ZTF are unresolved and may be from the host, and only early spectra exist -- these SNe may have developed broad lines at a later epoch, but are not included. We also do not include objects where only pre-peak spectra exist and the narrow SN lines could in principle be replaced by broad lines at a later epoch -- of these there are 12.}

The majority of the $gri$-band photometry in our study is from the ZTF Observing System \citep{dekany20} on the 48-inch Samuel Oschin Telescope in Palomar. This includes data from the public survey with a 3-day cadence and the ZTF partnership and Caltech surveys with a faster cadence ($\leq 2$\,days) over smaller areas \citep{bellm19}. Additional photometry was obtained using SEDM and the optical imager (IO:O) on the LT. {The ZTF data reduction and pipelines} are managed by the Infrared Processing and Analysis Center (IPAC) at Caltech as described by \citet{masci19}. We have also obtained photometry, including upper limits, from the IPAC forced photometry service\footnote{\url{http://web.ipac.caltech.edu/staff/fmasci/ztf/forcedphot.pdf}}. The IPAC photometry makes use of the ZOGY algorithm \citep{zackay16} for subtraction of reference images. The SEDM imaging data were processed using PSF photometry, and the magnitudes were calibrated against either SDSS or Pan-STARRS1 reference images, using \texttt{Fpipe} \citep{Fremling2016}. LT data were processed by a similar custom-built software \citep{Taggart2020}. ZTF and LT $griz$ magnitudes are calibrated to the Pan-STARRS1 photometric system 
and $u$-band data to the SDSS system. We include a table of all photometry used in this study as supplementary material; a sample of this table is presented in Table \ref{apptab:Phot}.

Reductions of P200/DBSP spectra were carried out using two pieces of software, \texttt{pyraf-dbsp} \citep{bellm2016} as well as a newer Python pipeline, \texttt{DBSP\_DRP} \citep{roberson21}, based on \texttt{PypeIt} \citep{prochaska20}. Keck/LRIS data reduction used the custom-written, publicly available IDL-based software, {\tt lpipe} \citep{perley19}. ALFOSC and ISIS spectra were reduced using standard procedures in \texttt{IRAF} \citep{tody86}. SPRAT spectra were reduced by the automated LT pipeline \citep[][]{barnsley12}. SEDM spectra were processed by the Python-based \texttt{pysedm} pipeline \citep{rigault19}; we note that the extracted spectra contain host galaxy light. All spectra presented in this paper will be available on WISEREP\footnote{\url{https://www.wiserep.org/}} \citep{yaron12}.

\emph{Swift}/UVOT data (for eight of the 14 SNe) were retrieved from the NASA Swift Data Archive\footnote{ \url{https://heasarc.gsfc.nasa.gov/cgi-bin/W3Browse/swift.pl}} and processed using the UVOT data analysis software \texttt{HEASoft} 6.19\footnote{ \url{https://heasarc.gsfc.nasa.gov/}}. The count rates were obtained from the images using the \emph{Swift} tool {\tt uvotsource}{, using a circular $3\arcsec$-radius region. The background was estimated using a significantly larger region close to the SN position. Counts} were converted to magnitudes using the UVOT photometric zero points \citep{breeveld11} and calibration files from September 2020. {Any host galaxies significantly contributing to the measured photometry\footnote{This applies to SNe 2019cqc, 2019xfs, 2020jhm and 2020yue.} (based on visual inspection of \emph{Swift} and \emph{GALEX} images of the SN and its location)} were subtracted from the \emph{Swift} data using template images taken in late 2021. We also checked for \emph{Swift} X-Ray Telescope (XRT) detections: no SN in the sample was detected in 0.2--10~keV XRT observations over a range of epochs from before the optical peak to $>2$ years after the peak. We list the XRT upper limits in Table \ref{apptab:XRT} in Appendix \ref{app:Xray}, where we also describe the process of estimating the limits.

\section{Spectroscopic evolution}
\label{sec:spec}

\begin{figure*}
\begin{minipage}{0.92\linewidth}
\centering
\includegraphics[width=\columnwidth]{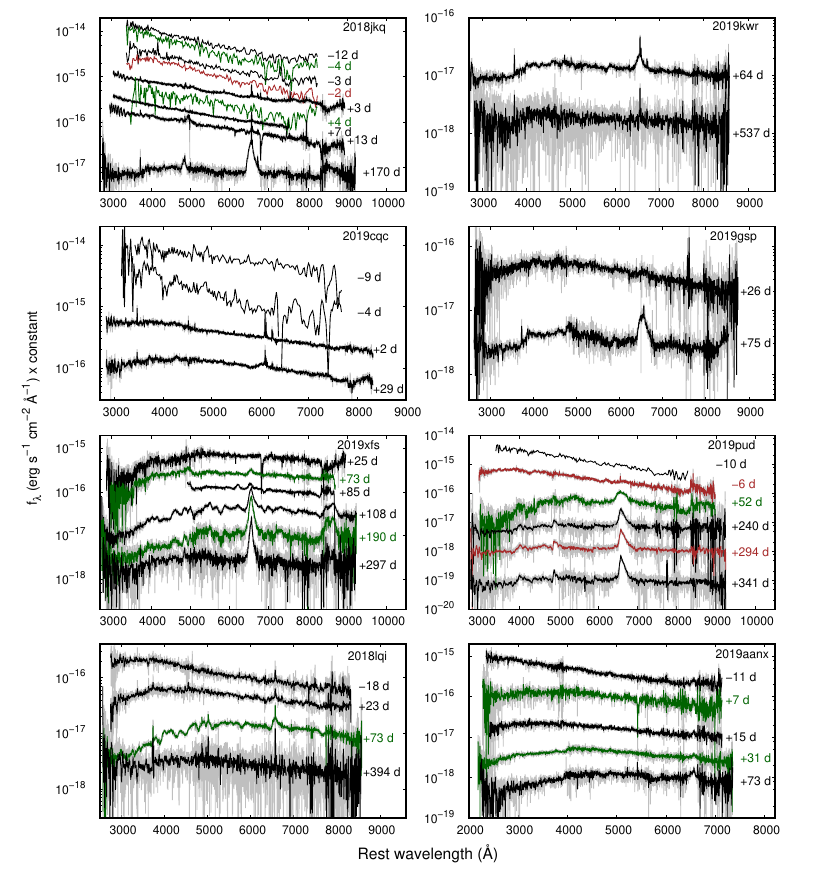}
\caption{Spectral sequences of the SNe in our sample. The colours of overlapping spectra are different for clarity. All epochs refer to rest-frame $g$-band light curve peaks. Savitzky-Golay smoothing has been applied; the original spectra are plotted in grey.}
\label{fig:spec_all1}
\end{minipage}
\end{figure*}

\setcounter{figure}{1}

\begin{figure*}
\begin{minipage}{0.92\linewidth}
\centering
\includegraphics[width=\columnwidth]{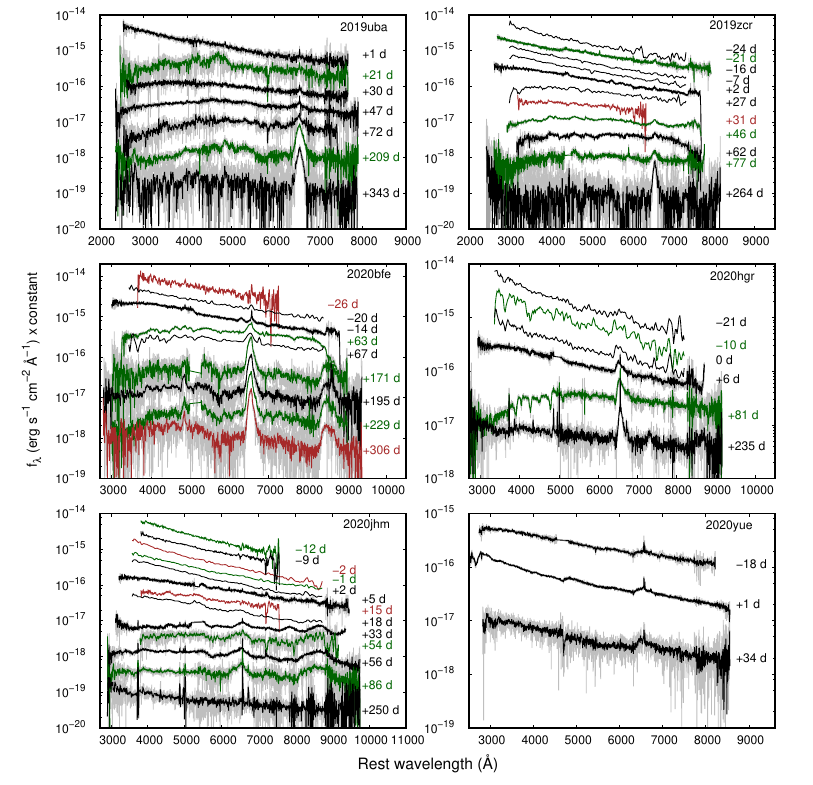}
\caption{-- continued.}
\label{fig:spec_all2}
\end{minipage}
\end{figure*}

\begin{figure*}
\begin{minipage}{0.85\linewidth}
\centering
\includegraphics[width=\columnwidth]{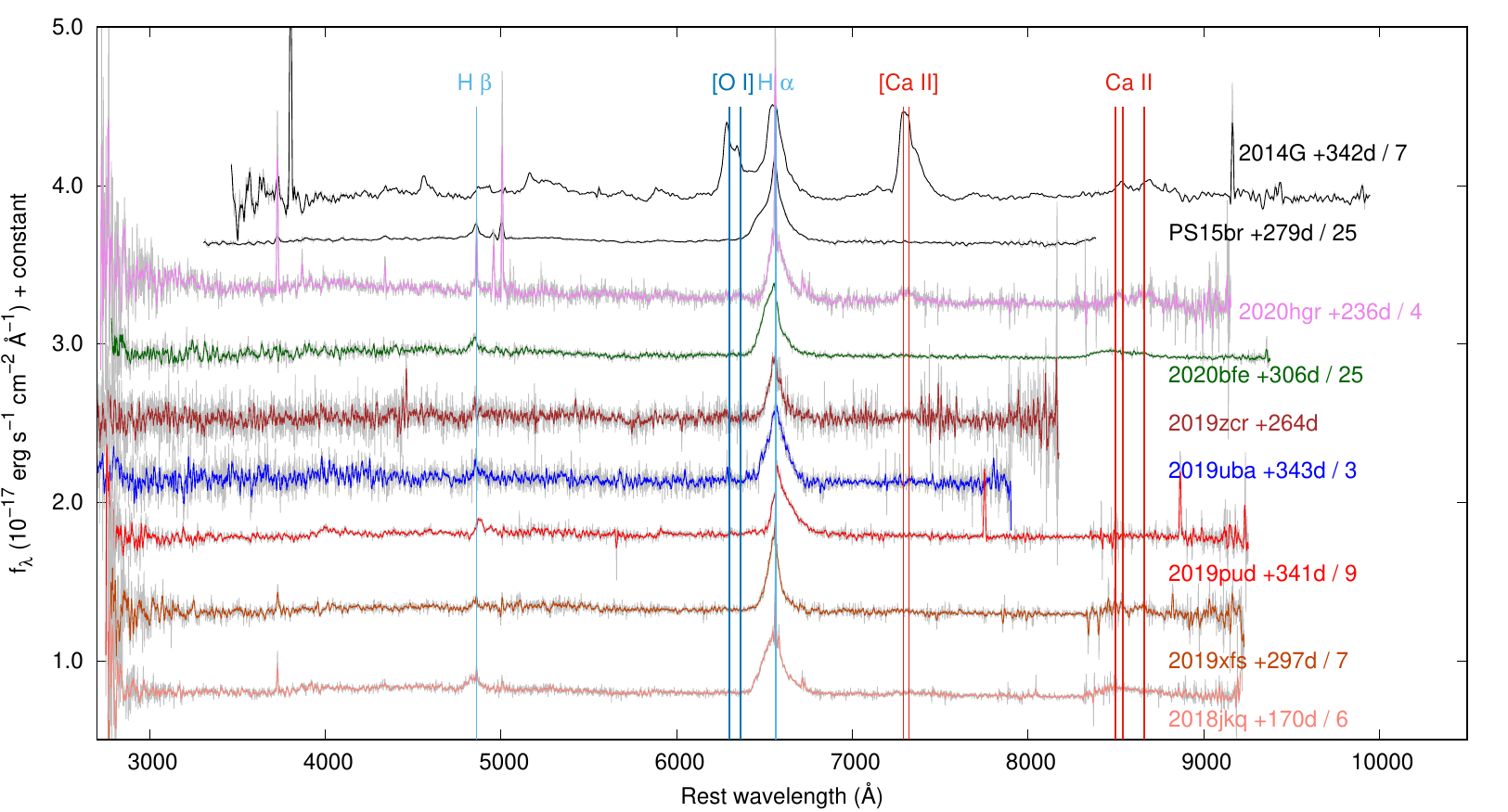}
\caption{Comparison between late-time spectra of the Type~II-L SN~2014G \citep{terreran16}, the SLSN II PS15br \citep{inserra18} and the seven SNe in our sample with available spectra in this phase, corrected for Galactic extinction.}
\label{fig:spec_neb}
\end{minipage}
\end{figure*}

\begin{figure*}
\begin{minipage}{0.85\linewidth}
\centering
\includegraphics[width=\columnwidth]{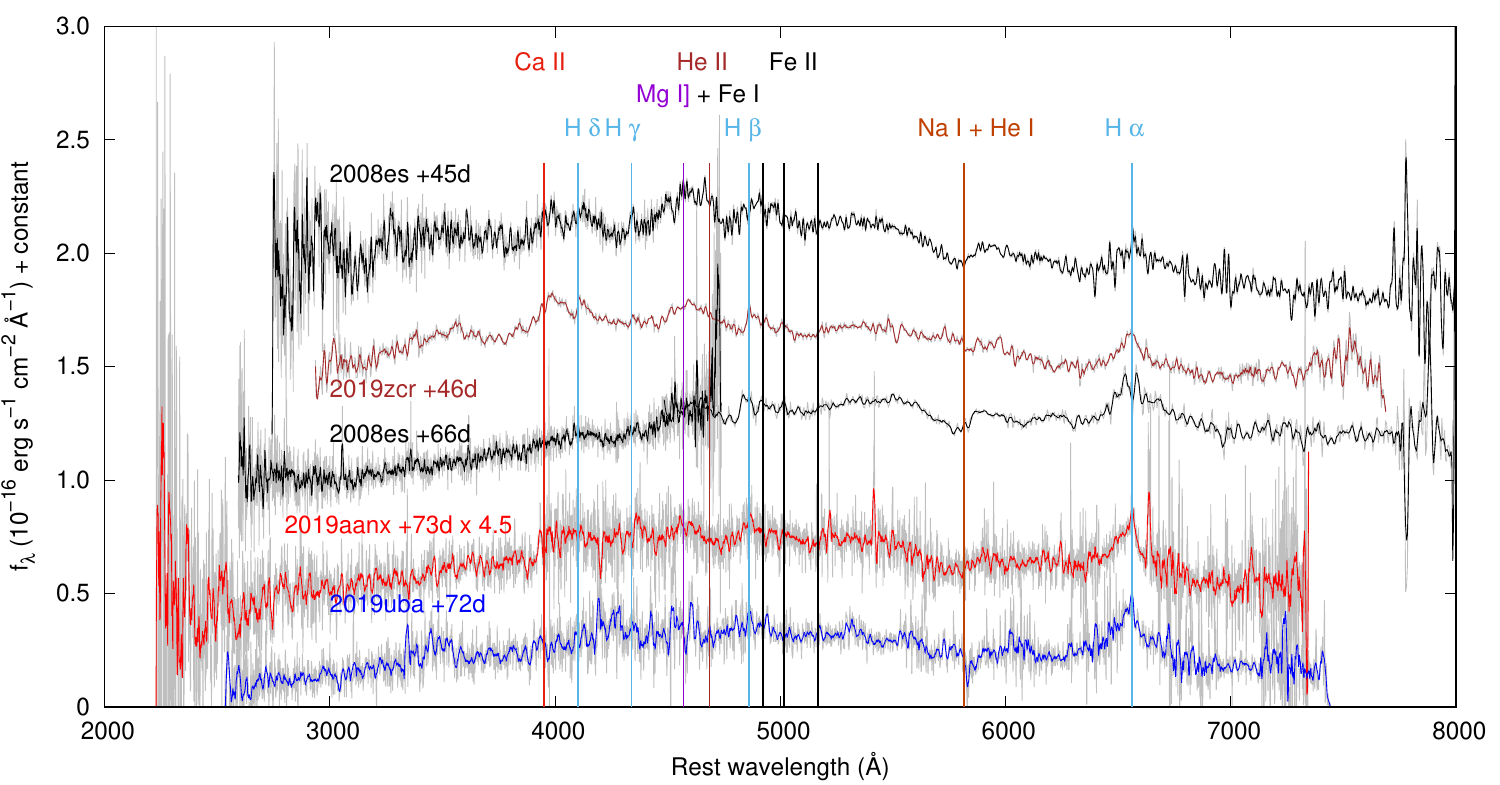}
\caption{Comparison between the spectra of SN 2008es \citep{gezari09, miller09} and the three similar SNe in our sample, corrected for Galactic extinction. }
\label{fig:spec_08es}
\end{minipage}
\end{figure*}

\begin{figure*}
\begin{minipage}{\linewidth}
\centering
\includegraphics[width=0.9\columnwidth]{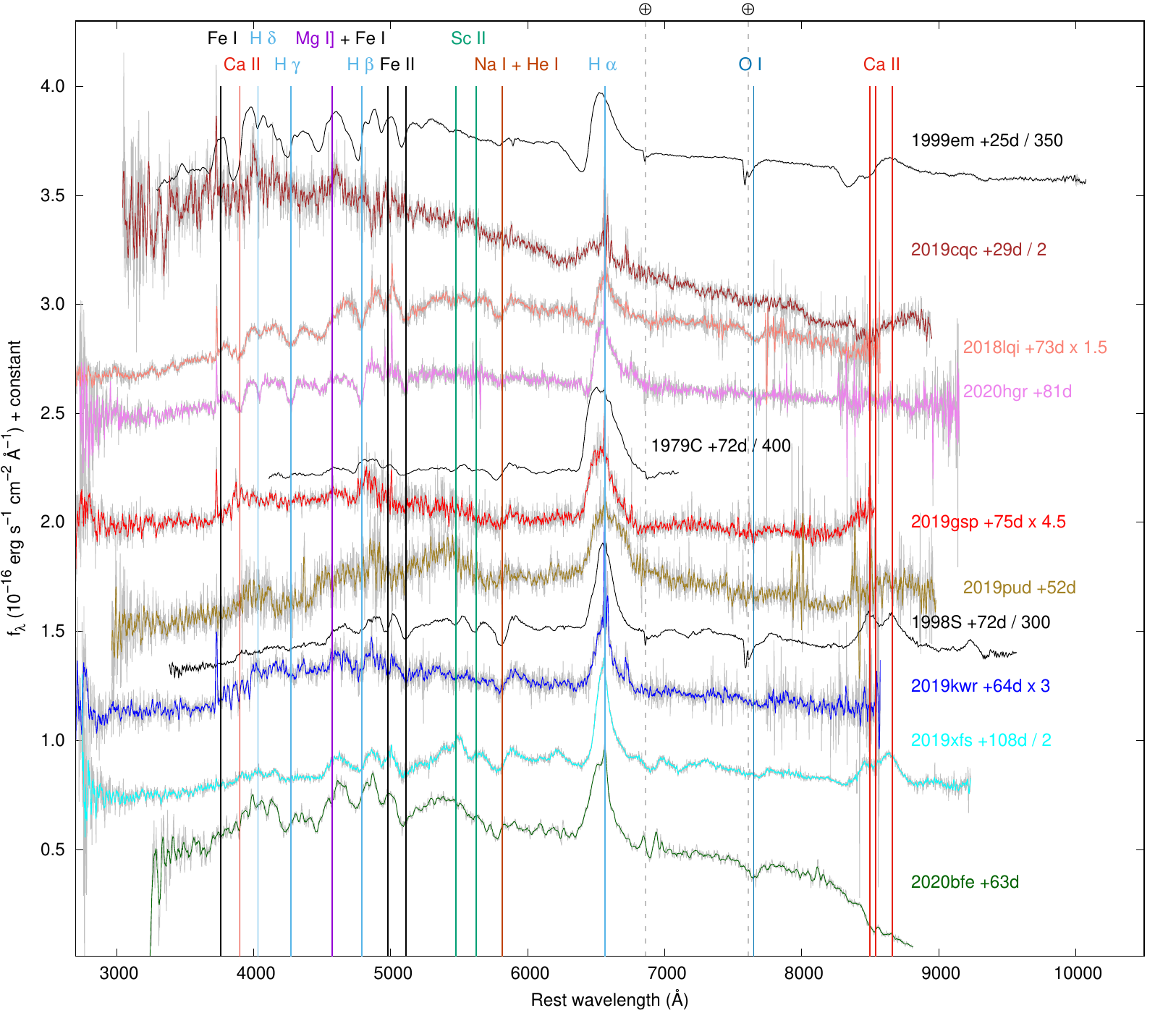}
\caption{Comparison between the spectra of normal Type II-P SN 1999em \citep{elmhamdi03}, the luminous Type II-L SNe 1979C \citep{branch91} and 1998S \citep{fassia01}, and similar SNe in our sample, corrected for Galactic extinction. Telluric lines ($\bigoplus$) have been marked for SNe 1999em, 1979C and 1998S; these have been removed from the ZTF spectra. }
\label{fig:spec_98s}
\end{minipage}
\end{figure*}

\begin{figure*}
\begin{minipage}{\linewidth}
\centering
\includegraphics[width=0.9\columnwidth]{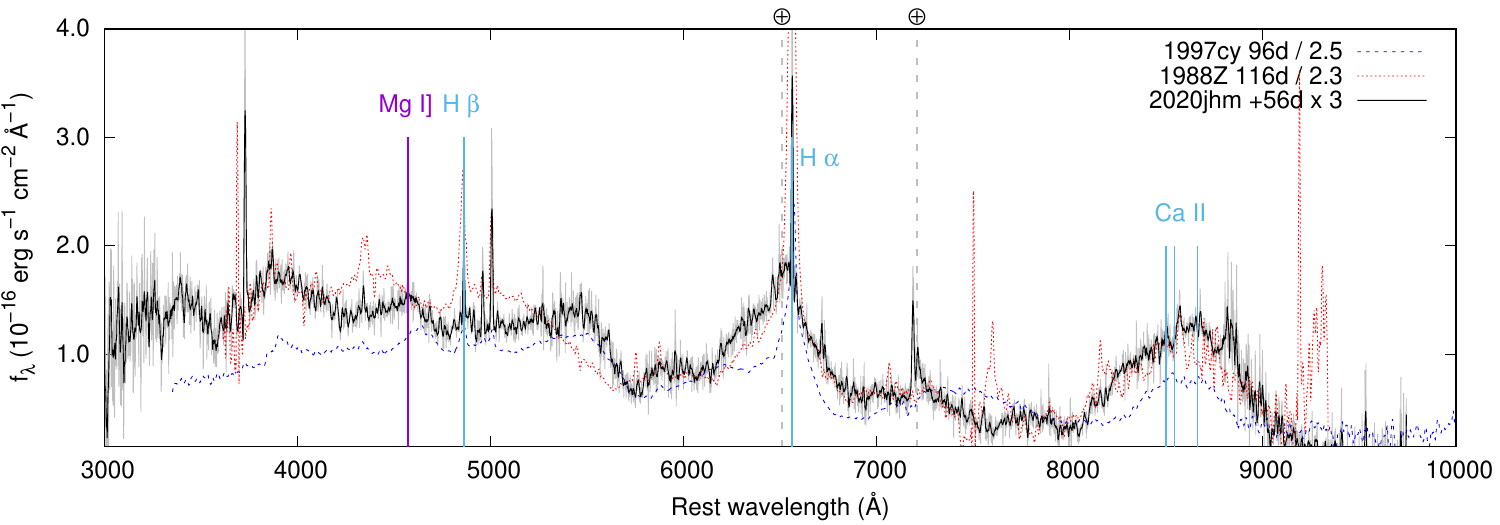}
\caption{Comparison between the spectra of SN~2020jhm (black solid) and interacting SNe 1988Z \citep[red dotted, Type IIn;][]{turatto93} and 1997cy {\citep[blue dashed, Type IIn or Ia-CSM;][]{turatto00,hamuy03}}, corrected for Galactic extinction. Similarities can be seen between all three events, although the evolution of SN~2020jhm is somewhat faster. Dates for SNe 1988Z and 1997cy refer to the discovery date. Telluric features ($\bigoplus$) are marked.}
\label{fig:spec_prqz}
\end{minipage}
\end{figure*}

\subsection{Common spectral features of the sample} 

{Our sample has been selected so that the only common factors are the luminosity and the presence of broad Balmer emission lines without strong narrow lines. A large fraction} of the sample nonetheless exhibits a broadly similar spectral evolution: an early phase with blue and nearly featureless spectra at and around the light curve peak, followed by the appearance of broad Balmer emission, sometimes with a P Cygni profile. The spectroscopic evolution is shown in Fig. \ref{fig:spec_all1} for each object in the sample. All phases below refer to the light curve peak in the rest-frame $g$ band (see Sect. \ref{sec:lc}). The broad emission lines are often contemporaneous with the appearance of absorption lines of Na~{\sc i} + He~{\sc i}, Fe~{\sc ii}, Sc~{\sc ii} and emission from Mg~{\sc i}] and Ca~{\sc ii}. The lines and their velocities {(5000--15000~km~s$^{-1}$)} in most of our objects resemble Type~II SNe in general, with weaker H$\alpha$ absorption than is typical for SNe~II-P but similar to SNe~II-L. The prototypical SN~2008es has also been shown to bear some similarity to the luminous Type II-L SN~1979C \citep{gezari09,miller09}, albeit with a delayed spectroscopic evolution, as did CSS121015 \citep{benetti14} and PS15br \citep{inserra18}.

A strong broad H$\alpha$ line mostly does not develop until weeks after maximum light, but a weak emission is seen in the early spectra of some SNe in the sample. {Late-emerging} Balmer lines are often seen in SNe~II-L {\citep[e.g.][]{fassia01,kangas16,terreran16}}, and also seen in three of the four previously observed {SLSNe~II}: SN~2008es, CSS121015 and PS15br. However, this is not ubiquitous: {SNe~2020hgr and 2020yue develop a strong broad emission line before the peak, as did SN 2013hx \citep{inserra18} and the normal Type II-L SN 1980K \citep{uomoto86}.}

We have late-time ($>200$~d) spectra available for six of the 14 SNe in our sample\footnote{{This does not include late-time spectra (those of SNe~2019kwr, 2018lqi and 2020jhm) where only host galaxy lines are seen.}}. {Additionally, for SN~2018jkq, we have a +170~d spectrum}. The latest-phase spectra of these SNe are shown in Fig. \ref{fig:spec_neb}, together with a comparable-epoch spectrum of PS15br \citep{inserra18} for comparison. The spectrum is generally dominated by a strong, broad H$\alpha$ emission line (FWHM $\gtrsim5000$~km~s$^{-1}$), accompanied by a weaker H$\beta$ line and little else. 
When its wavelength is covered, we see the Ca~{\sc ii} infrared triplet except in the case of SN~2019pud, where the comparatively weak blue wing of the H$\alpha$ profile is also exceptional. 

Similarly to PS15br and SN 2013hx \citep{inserra18}, we see no clear [O~{\sc i}]~$\lambda\lambda6300,6364$ emission, which is typically strong in normal SNe~II \citep[e.g.][]{terreran16,dessart20} and SLSNe~I {\citep[e.g.][]{chen15,jerkstrand17}} at similar epochs. The typically similarly strong [Ca~{\sc ii}]~$\lambda\lambda7291,7323$ doublet is weak or nonexistent as well, only clearly seen in SN~2020hgr. There is a difference as well: {PS15br and SN~2013hx} evolved to show a multi-component H$\alpha$ profile at $\gtrsim230$~d, suggesting late-time CSI. In our spectra, the emission profile typically only shows one component. A possible narrower component is visible in SN~2019pud at $\gtrsim300$~d and SN~2020bfe at $\gtrsim190$~d, {but the multi-component profile} is unclear compared to \citet{inserra18}. 

\subsection{Spectroscopic sub-groups}

There are some indications of sub-groups within the sample. Three SNe in the sample are quite similar to SN~2008es in terms of their photospheric-phase spectra; these are SNe~2019aanx, 2019uba and 2019zcr, and we henceforth refer to these SNe as ``08es-like". The same SNe are also similar to SN 2008es photometrically (see Sect. \ref{sec:lc}). This resemblance is demonstrated in Fig. \ref{fig:spec_08es}. Not all features of the SN~2008es spectra are replicated, but this is at least partially due to the noisy spectra of these distant ($z=0.26$ to $z=0.40$) SNe. We do, however, see similar shapes (broad, symmetric emission with no P Cygni absorption) and widths (6000--8000 km s$^{-1}$) in their H$\alpha$ profiles, and weak absorption lines compared to the rest of the sample, {at the same phases.} The broad component shows a possibly Lorentzian profile, albeit without the strong narrow core typical for SNe~IIn. {An intermediate or emerging/weak broad emission component is seen around peak}, similarly to PS15br \citep{inserra18}. Signs of a ``bump" feature are also seen blueward of $\sim5700$~\AA, albeit quite weak in the available spectra of SN~2019uba. Such a feature is associated with a pseudo-continuum comprised of a forest of blended iron lines, often seen in interacting Type IIn and Ibn SNe \citep[e.g.][]{turatto93,karamehmetoglu21,kool21}.{ The symmetric shape of the H$\alpha$ emission profile continues until late times. }

A majority of our sample, however, spectroscopically resembles typical SNe~II more than SN~2008es. These are SNe~2019kwr, 2019cqc, 2019gsp, 2019xfs, 2018lqi, 2020bfe, 2020hgr and 2020yue. We show a comparison between SNe 1999em, 1979C and 1998S and these SNe (with the exception of SN~2020yue as it lacks spectra in the shown phases) in Fig. \ref{fig:spec_98s}. We also include SN~2019pud in this figure despite its early-time peculiarity (see below). Features seen in these events include a selection of absorption and P Cygni lines typical to Type II SNe (O {\sc i}, Na {\sc i} + He {\sc i}, Sc {\sc ii}, Fe {\sc ii} and Mg {\sc i}]), with relatively shallow or often undetected absorption in H$\alpha$. These features are typical for Type II-L spectra in particular. The widths of the absorption lines extend to roughly $-10000$~km~s$^{-1}$ in the phases ($>+50$~d) where they are clearly visible. 
We only have early spectra of SN~2020yue, which resemble those of a young SN II-P with strong, broad P Cygni profiles for the Balmer lines. This is atypical for our sample, where early spectra tend to be featureless. For SN~2019cqc our latest-phase spectrum is at +29~days, which shows a shallow but very broad P Cygni feature (extending up to $\sim-20000$~km~s$^{-1}$) in H$\alpha$ and in the Ca {\sc ii} NIR triplet. These lines are likely still in the process of emerging, as the peak-phase spectrum is featureless.

The last sub-group consists of three SNe with more peculiar spectra or uncertain evolution. SN~2018jkq was not spectroscopically observed between the peak (featureless apart from weak, narrow Balmer lines) and the late phase, but the spectrum at +170~d exhibits broad (FWHM $\sim5500$~km~s$^{-1}$) emission of H$\alpha$, H$\beta$ and Ca {\sc ii} very similarly to several other SNe in our sample. SN~2019pud eventually starts to resemble SN~1979C, but early on shows a broad absorption feature at $\sim4250$~\AA, atypical for Type II SNe. Multiple lines could be responsible for this feature, such as those of Fe~{\sc iii} or O~{\sc ii}; we cannot definitively identify this line. Finally, {SN~2020jhm is spectroscopically unique within the sample}. Its early spectra are blue and featureless, but around +30~d it develops strong, broad emission lines of H$\alpha$, the Ca {\sc ii} NIR triplet and O {\sc i}~$\lambda7774$. The O {\sc i} emission then rapidly weakens, while the other emission lines eventually extend to $\sim-18000$~km~s$^{-1}$. The pseudo-continuum of iron lines around 5600~\AA \ is also seen in this SN. Apart from the absence of {narrow} emission lines, the spectrum resembles SN 1988Z \citep{turatto93}, albeit at an earlier phase; see Fig.~\ref{fig:spec_prqz}. {SNe~2019pud and 2020jhm also exhibit peculiar early light curves} {(see Sect. \ref{sec:lc})}.

\section{Light curves}
\label{sec:lc}

\subsection{Peak fits and absolute magnitudes}
\label{sec:lccorrs}

{We have performed} numerical interpolation of each light curve using a Gaussian process (GP) regression algorithm \citep{rasmussen06} {to obtain peak magnitudes and epochs.} We used the Python-based {\tt george} package \citep{hodlr}, which implements various different kernel functions. {Mat{\'e}rn kernels with $\nu$ parameter of 3/2 or 5/2 were used.} The uncertainty in peak epoch is estimated as the time range when the GP light curve is brighter than the $1\sigma$ lower bound on the peak brightness.

\begin{figure}
\centering
\includegraphics[width=0.9\columnwidth]{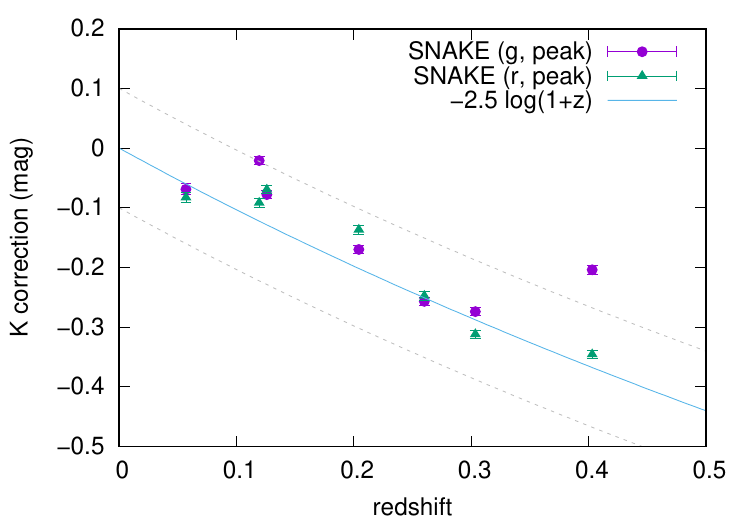}
\caption{Approximate $K$-corrections using $- 2.5\mathrm{log}(1 + z)$ (line) compared to the values determined using \texttt{SNAKE} \citep{inserra18} from spectra close to peak for the SNe where this was possible (points). The dashed lines correspond to $\pm0.1$ mag, and contain most of the variation.}
\label{fig:kcorrs}
\end{figure}

$K$-corrections \citep{hogg02} were performed using the $- 2.5\, \mathrm{log}(1 + z)$ approximation. For the SNe where we have a spectrum close to the light curve peak (7 out of 14 events), we have also determined the spectroscopic $K$-correction using the \texttt{SNAKE} code \citep{inserra18}. As seen in Fig. \ref{fig:kcorrs}, for these SNe the approximation is good to within $\sim0.1$ mag. A similar trend was shown for SLSNe~I by \citet{chen22a}. Spectroscopic $K$-corrections are not possible for the majority of epochs, and we have thus, for the sake of uniformity, used the approximation for all photometric points. Thus we have obtained the absolute light curves of each SN using
\begin{equation}
    M_{\mathrm{R}} = m_{\mathrm{O}} - \mu - A_{\mathrm{O,Gal}} + 2.5\, \mathrm{log}(1 + z),
\end{equation}
where R stands for the rest-frame filter, O is the observed filter, $\mu$ is the distance modulus, and $A_{\mathrm{O,Gal}}$ is the Milky Way extinction \citep{schlafly11} in filter O determined using the \citet{cardelli89} law. For redshifts $z<0.17$ (7 out of 14 events), filters O and R are the same. Otherwise R is the filter with the closest effective wavelength to the redshift-corrected effective wavelength of O, in the sequence of UVW2, UVM2, UVW1, $u$, $g$, $r$, $i$ and $z$.

\begin{figure*}
\begin{minipage}{\linewidth}
\centering
\includegraphics[width=0.9\columnwidth]{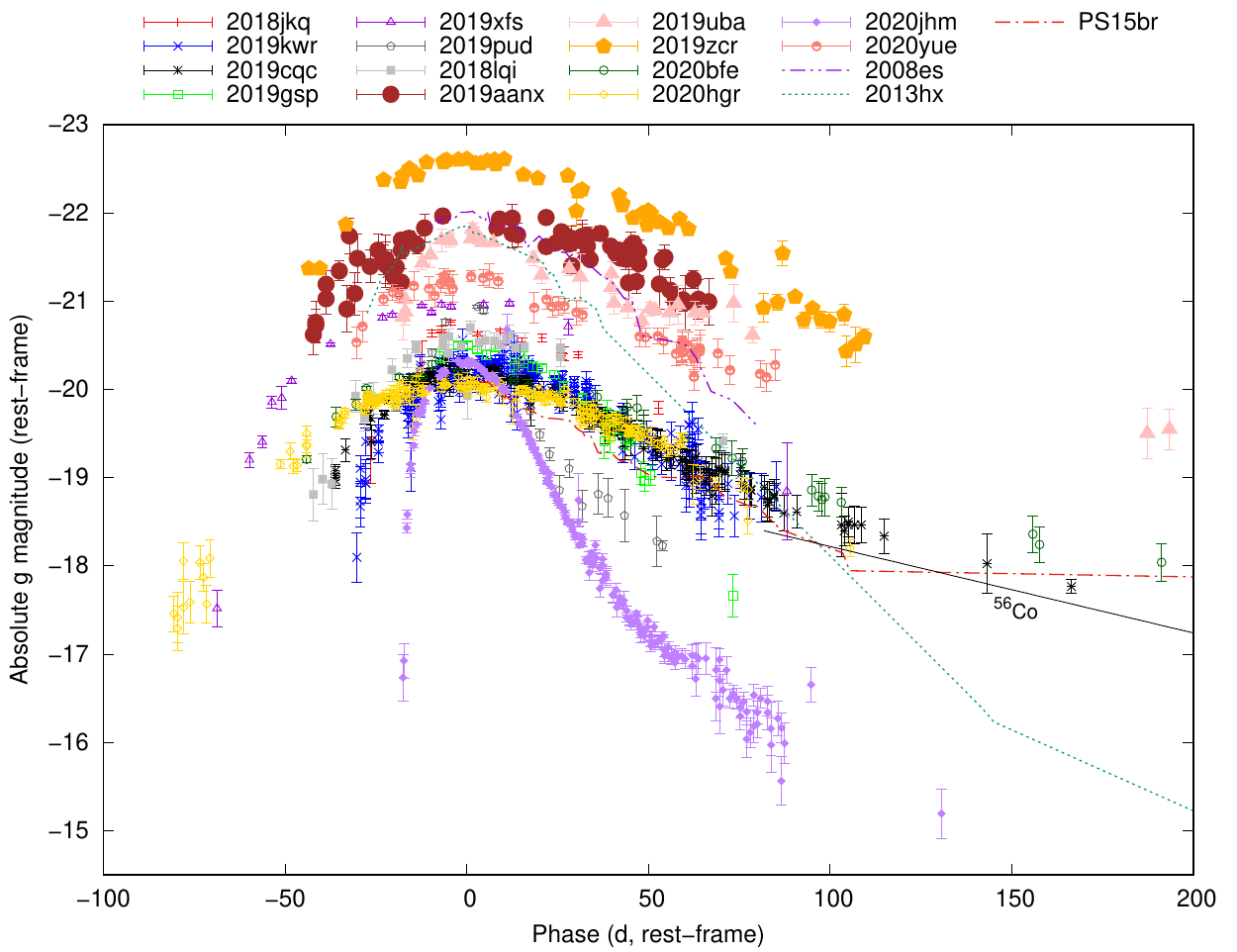}
\caption{Absolute-magnitude light curves of our sample SNe in the rest-frame $g$ band (points). We also show the previously studied SLSNe~II: SN 2008es, SN 2013hx and PS15br \citep[lines;][]{gezari09,inserra18}. {Large symbols correspond to the 08es-like SNe in our sample.} The phase refers to rest-frame days after the light curve maximum. The heterogeneity of the SNe is apparent in terms of peak magnitude and decline rate.}
\label{fig:lc_g}
\end{minipage}
\end{figure*}

\begin{figure*}
\begin{minipage}{0.9\linewidth}
\includegraphics[width=\columnwidth]{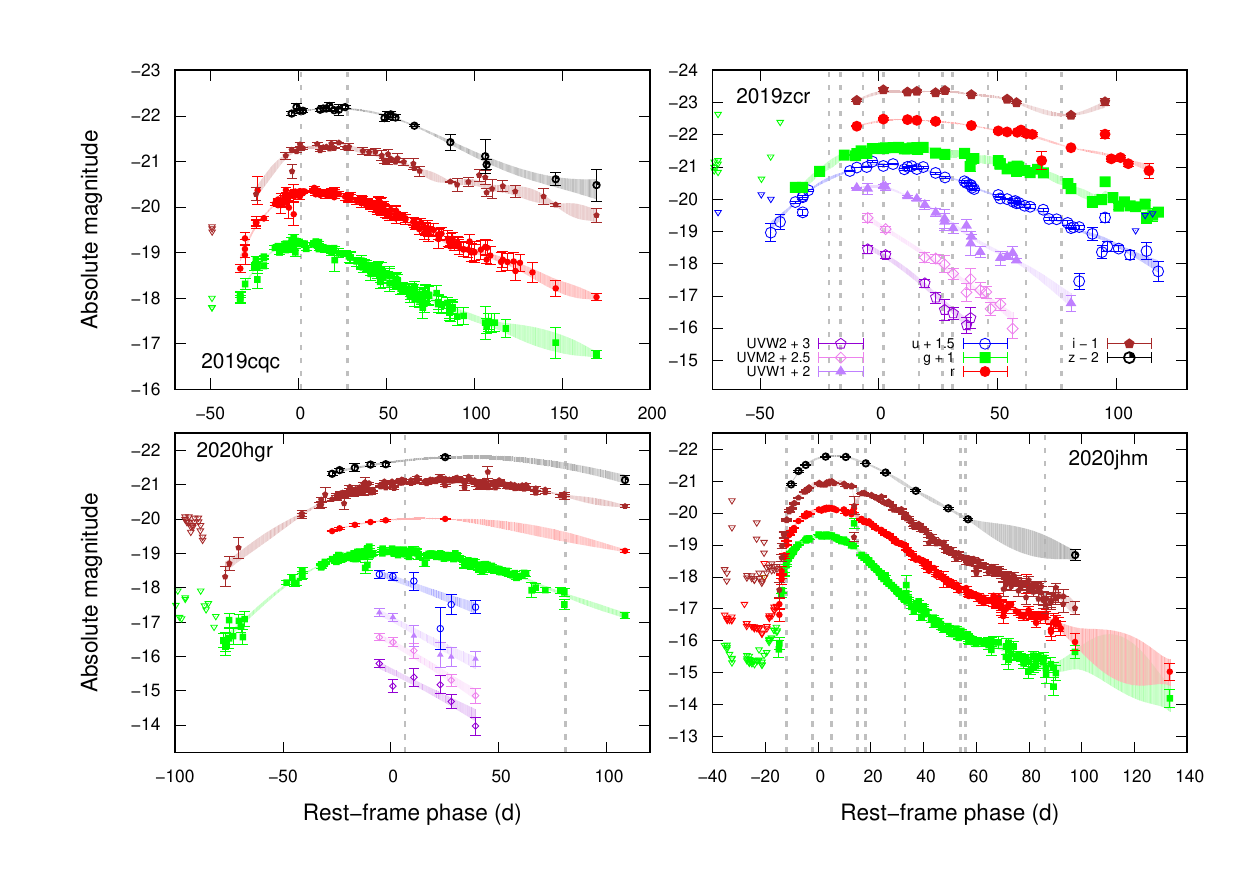}
\caption{Our GP fits to the light curves of the four SNe in our sample with {sufficient coverage in at least four bands. Large symbols correspond to an 08es-like SN.} Vertical dashed lines correspond to the epochs of spectra.}
\label{fig:george2}
\end{minipage}
\end{figure*}

\begin{figure*}
\begin{minipage}{0.9\linewidth}
\centering
\includegraphics[width=\columnwidth]{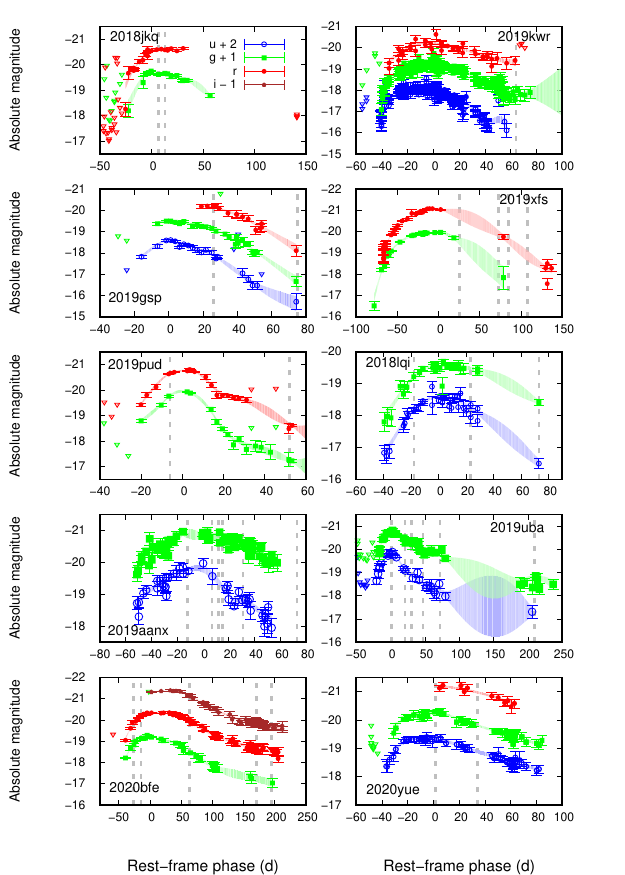}
\caption{Our GP fits to the light curves of the ten SNe in our sample {with sufficient coverage in only two or three bands}. {Large symbols correspond to the 08es-like SNe.} Vertical dashed lines correspond to the epochs of spectra.}
\label{fig:george1}
\end{minipage}
\end{figure*}

The peak parameters, rise times from half-maximum and decline parameters of our SNe are listed in Table \ref{tab:george}. Here we use both $\Delta g_{50}$, i.e. the decline from the $g$-band light curve peak in 50 rest-frame days, and $s_2$, the decline rate during the late photospheric phase \citep{anderson14}. We show the absolute rest-frame $g$ band light curves of all sample SNe together in Fig. \ref{fig:lc_g}. The individual multi-band GP light curve of each SN is shown in Figs. \ref{fig:george2} {(SNe with GP fits in at least four bands)} and \ref{fig:george1} {(the rest)} for the filters where a GP fit is feasible. We have used the GP fits to obtain the rest-frame $g-r$ colours when possible; we show these in Fig. \ref{fig:georgeC}.

\begin{table*}
\begin{minipage}{0.75\linewidth}
    {\centering
    \caption{Rest-frame $g$-band peak absolute magnitudes, peak epochs, rise times from half-maximum ($t_\mathrm{rise,0.5}$) and decline parameters $\Delta g_{50}$ and $s_2$ of our sample SNe based on GP fits using {\tt george} \citep{hodlr}. The peak magnitudes are de-reddened for Galactic extinction and $K$-corrected. The errors quoted here are solely based on the GP fits.}
    \begin{tabular}{lccccc}
       SN & Peak MJD & Abs. $g$-band peak & $t_\mathrm{rise,0.5}$ & $\Delta g_{50}$ & $s_2$ \\
       & (d) & (mag) & (d, rest) & (mag) & [mag (100 d)$^{-1}$] \\
        \hline
        SN~2018jkq & $58475.5^{+2.5}_{-2.7}$ & $-20.74\pm0.04$ & $15.5^{+2.4}_{-2.5}$ & $0.70\pm0.14$ & $1.6\pm0.3$ \\
        SN~2019kwr & $58589.0\pm3.8$ & $-20.25\pm0.03$ & $32.7^{+3.4}_{-3.2}$ & $1.00\pm0.05$ & $2.7\pm0.2$ \\ 
        SN~2019cqc & $58595.2^{+4.1}_{-3.7}$ & $-20.21\pm0.02$ & $27.7^{+3.8}_{-3.4}$ & $0.72\pm0.04$ & $1.3\pm0.1$ \\
        SN~2019gsp & $58636.1\pm6.7$ & $-20.54\pm0.03$ & $15.9\pm5.8^{a}$ & $1.36\pm0.06$ & $4.5\pm0.6$ \\ 
        SN~2019xfs & $58809.6^{+6.4}_{-15.1}$ & $-20.97\pm0.03$ & $55.3^{+5.8}_{-13.6}$ & $1.14\pm0.50$ & $2.6\pm0.3$ \\ 
        SN~2019pud & $58754.6\pm3.6$ & $-20.96\pm0.05$ & $13.7^{+3.3}_{-3.4}$ & $2.63\pm0.12$ & $3.2\pm0.3$ \\ 
        SN~2018lqi & $58783.7^{+13.1}_{-10.9}$ & $-20.57\pm0.04$ & $30.3^{+11.0}_{-9.2}$ & $0.67\pm0.22$ & $2.3\pm0.2$  \\ 
        SN~2019aanx & $58828.4^{+1.0}_{-22.4}$ & $-21.92\pm0.02$ & $46.0^{+0.8}_{-16.0}$ & $0.82\pm0.10$ & $2.1\pm0.4$ \\  
        SN~2019uba & $58810.2\pm4.0$ & $-21.70\pm0.02$ & $>18.3^{b}$ & $0.81\pm0.04$ & $1.3\pm0.3$  \\ 
        SN~2019zcr & $58901.4^{+1.6}_{-7.8}$ & $-22.61\pm0.07$ & $33.4^{+1.8}_{-6.5}$ & $0.61\pm0.08$ & $2.4\pm0.2$  \\ 
        SN~2020bfe & $58919.5^{+7.3}_{-6.3}$ & $-20.20\pm0.02$ & $35.3^{+6.7}_{-5.8}$ & $0.44\pm0.06$ & $1.9\pm0.2$  \\ 
        SN~2020hgr & $58990.7^{+5.1}_{-4.2}$ & $-20.06\pm0.01$ & $42.9^{+4.6}_{-3.8}$ & $0.55\pm0.02$ & $2.2\pm0.2$  \\ 
        SN~2020jhm & $58990.9^{+3.5}_{-0.7}$ & $-20.33\pm0.03$ & $11.3^{+3.4}_{-0.8}$ & $2.95\pm0.04$ & $2.6\pm0.4$ \\ 
        SN~2020yue & $59193.3^{+8.1}_{-7.6}$ & $-21.26\pm0.03$ & $>30.9^{b}$ & $0.68\pm0.04$ & $1.3\pm0.2$  \\
        \hline
    \end{tabular}
    \label{tab:george}}
    $^{a}$The rise of SN~2019gsp is only covered by the rest-frame $B$-band light curve, which is used here. \\
    $^{b}$The rises of SNe~2019uba and 2020yue are not fully covered by our observations, and a lower limit based on the first detection is given. 
    \end{minipage}
\end{table*}  

Any host galaxy extinction has not been accounted for. We do not detect narrow Na {\sc i} D absorption in any of our spectra, indicating that host galaxy extinction is {moderate at worst}. We note that the peak-phase rest-frame $g-r$ colours within the sample vary from roughly 0.2 to $-0.2$ mag (see below and Fig. \ref{fig:georgeC}), which may indicate differences {in host galaxy extinction.} There is, however, considerable diversity within the sample in terms of light curves and spectra, and intrinsic colour differences are likely to exist. Therefore, we have not attempted to use colour differences to determine host extinction. 

\subsection{Photometric properties}
\label{sec:lcprop}

From Fig. \ref{fig:lc_g}, it is clear that the light curves of the class formed by our sample and the SLSNe II previously studied by \citet{inserra18} are somewhat heterogeneous. Most of the events are clustered between peak $g$-band magnitudes of roughly $-20$ and $-21$~mag{. This group includes a range of decline time scales, but most} of the light curves decline at a similar rate, between 1.5 and 2 mag (100 d)$^{-1}$. This also applies to the second group within our sample, at peak magnitudes brighter than $-21$~mag -- but not to SN 2013hx and SN 2008es, which {decline} at $\sim4$~mag (100 d)$^{-1}$. There is also considerable variation in rise times from half-maximum (Table \ref{tab:george}), which range from 10--20~d (SNe~2018jkq, 2019pud and 2020jhm) to $\sim55$~d (SN~2019aanx) in the rest frame, mostly clustering around 30--50~d. This heterogeneity is therefore not only seen between, but also within, the spectroscopic subgroups.

\begin{figure}
\centering
\includegraphics[width=\columnwidth]{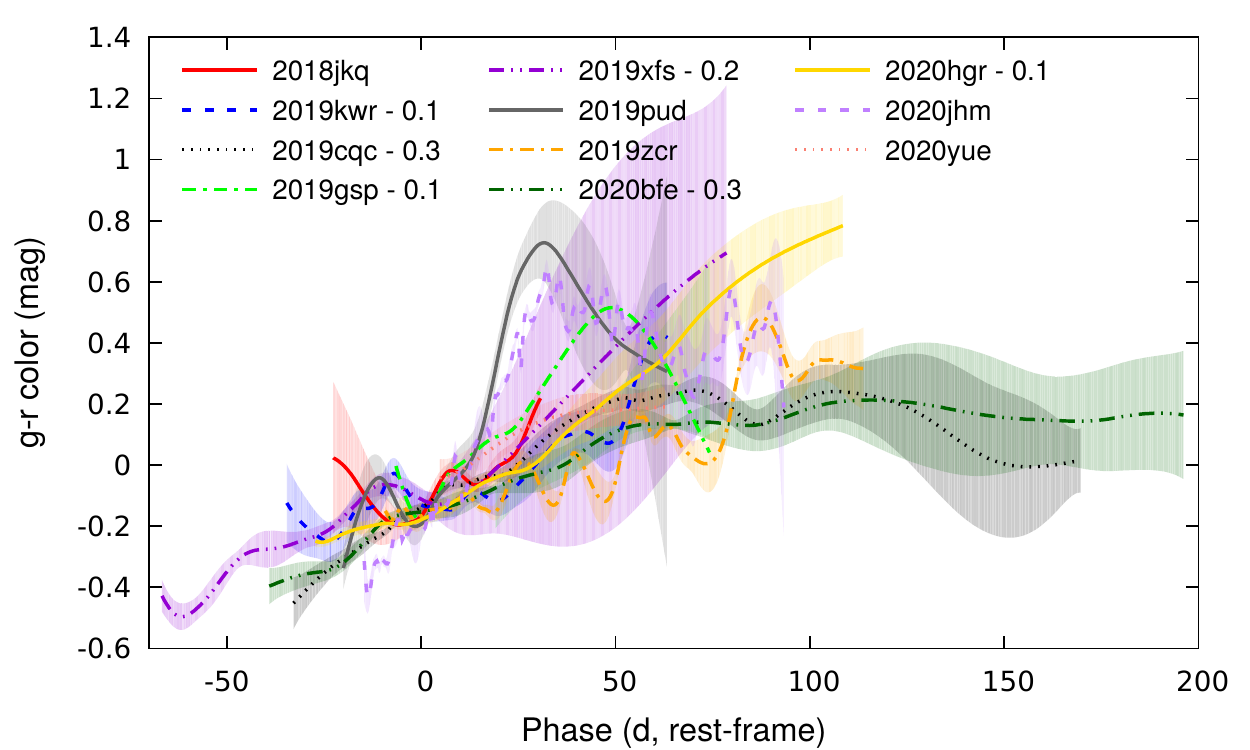} 
\caption{Our GP fits to the rest-frame $g-r$ colours of our sample objects when available. Constants have been added to roughly match peak-phase colours and emphasize the subsequent evolution. These constants may reflect intrinsic colour differences or low-to-moderate host galaxy extinction.}
\label{fig:georgeC}
\end{figure}

Some heterogeneity is also seen in the colour evolution. The fast-evolving events SN~2019pud and SN~2020jhm stand out in Fig. \ref{fig:georgeC}. From $\sim$20~d onwards (rest-frame), they redden quickly from $g-r\approx-0.2$~mag at the peak to $g-r \sim 0.6$~mag, then their evolution flattens or even moves back toward the blue from $\sim$35~d onwards. The slowest-evolving SN in the sample, SN~2020hgr, reddens until at least 110~d and reaches a colour of $g-r\gtrsim0.8$~mag{. The rest} of the sample tends to redden slower, reaching $g-r\sim0.5$~mag around 100~d. The colours at the peak lie roughly between $-0.2$ and 0.1 mag. Intrinsic colour differences are likely considering the variety in other features, but if this range was entirely due to host extinction, a difference of 0.3~mag in $g-r$ colour would correspond to $A_V \lesssim 1$~mag \citep{cardelli89}.

\begin{figure}
\centering
\includegraphics[width=0.98\columnwidth]{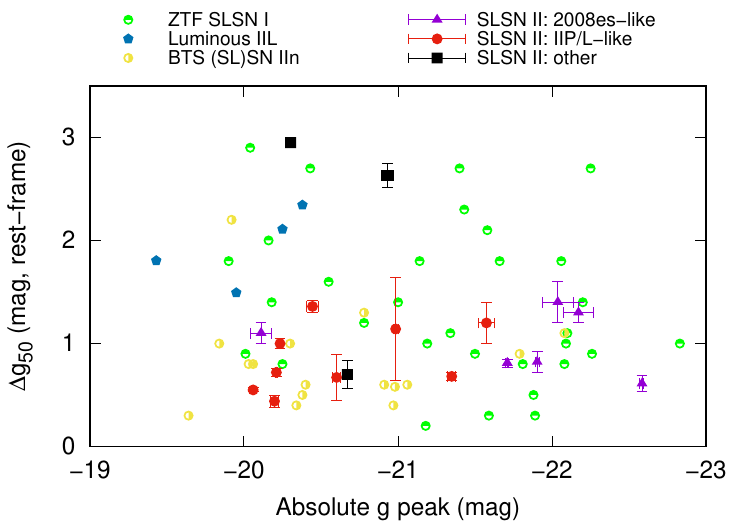}
\includegraphics[width=0.98\columnwidth]{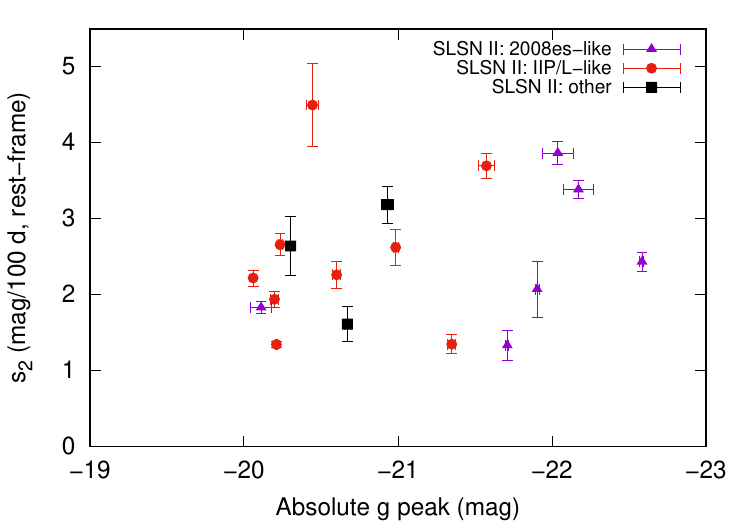}
\caption{Top: Decline rate characterized by the decline in 50 rest-frame days from the peak vs. the absolute (rest-frame) $g$-band peak magnitudes of the SNe in our sample, CSS121015 \citep{benetti14} {and the SNe of \citet{inserra18}}. We also include groups of other publicly available SNe for comparison \citep{devauc81,fassia00,kangas16,perley20,reynolds20,chen22a}. Bottom: decline rate $s_2$ \citep{anderson14} vs. $g$-band peak magnitude for the same SLSNe~II.}
\label{fig:peakdelta}
\end{figure}

To illustrate the range of light curve behaviour, we show the decline rate (see Table \ref{tab:george}) as a function of the peak magnitude in Fig. \ref{fig:peakdelta}. Here we measure the former with the quantity $\Delta g_{50}$, determined using GP interpolation. We have also measured the decline rates quantified as $s_2$ \citep{anderson14} for our sample. This is defined in normal SNe II as the decline rate in the plateau or bump phase, following an initial, steeper decay $s_1$. In most of our sample, we cannot clearly separate $s_1$ and $s_2${, in which case} we have measured $s_2$ after the light curve has clearly turned over from the peak.

In the top panel, we also include the SLSNe~II in \citet{inserra18}, CSS121015 \citep{benetti14} and the prototypical SLSN II, SN 2008es \citep{gezari09,miller09}. We also include various SNe of different types and separate our sample events into groups based on spectroscopic similarities (see Sect. \ref{sec:spec}). Here we consider CSS121015 and PS15br spectroscopically 08es-like, while the spectrum of SN 2013hx evolves to resemble normal SNe II \citep[see Fig. 2 in][]{inserra18}. {It is apparent that the brightest SLSNe~II are all 08es-like. The other SNe, most of which resemble normal Type II SNe more, tend to be fainter. However, PS15br is among the faintest in the sample and thus an outlier.}

Most of our sample is located below $\Delta g_{50} \approx 1.5$ mag, as are the previously studied SLSNe~II. There are two outliers: SNe~2019pud and 2020jhm (respectively, with $\Delta g_{50}$ of 2.6 and 3.0 mag). {In terms of $s_2$,} these SNe are not outliers, however; {their early steep decline} does not persist into late times. Both of these SNe were classified Type I.5, while the third SLSN I.5 in this sample, SN~2019zcr, is the brightest (peaking at $-22.61\pm0.07$ mag) and among the slowest ($\Delta g_{50} \approx 0.6$ mag). Spectroscopically, the latter resembles SN 2008es, while the former two do not closely resemble either SN 2008es, normal SNe~II, nor each other (see Sect. \ref{sec:spec}). Thus it is clear that the events classified as Type~I.5 do not form a unified group; the classification of both fast-evolving outliers as Type~I.5 may be a coincidence. The third SN in the ``other" group in Fig. \ref{fig:peakdelta}, SN~2018jkq, is in this group due to a lack of spectra, not any observed difference in evolution. It is photometrically similar to the ``II-P/L-like" group, though. 

Most of the light curves of this sample never exhibit a late-time decline similar to $^{56}$Co decay, i.e. 0.98~mag~(100~d)$^{-1}$; however, it is possible that most of the observed light curves simply do not reach that phase. Three SNe, SN~2019cqc, SN~2019uba and SN~2020bfe, have light curves extending to $>150$~d and do show a flattening decline. The rest-frame \emph{g}-band decline rates at $>100$~d in these events are $1.01\pm0.04$, $0.10\pm0.72$ and $0.74\pm0.04$~mag~(100~d)$^{-1}$ respectively; thus only SN~2019cqc is within 1$\sigma$ of the $^{56}$Co decay rate, while the large uncertainty of SN~2019uba puts it within 2$\sigma$. SN~2020bfe, meanwhile, declines too slowly for $^{56}$Co decay, which can be caused by a CSM or magnetar power source still being dominant at late times. 

\subsection{Comparison to other events} 
\label{sec:lccomp}

The rise times from half-maximum within our sample can mostly be determined from the {GP fit}. {The median rise time} is $\tilde{t}_\mathrm{rise,0.5} = 33\pm6$~d, and three SNe (SNe~2019xfs, 2019aanx and 2020hgr) have $t_\mathrm{rise,0.5}>40$~d. This is not only {much} longer than is typical for Type II SNe \citep[where the rise times from \emph{explosion} have a $1\sigma$ range of 4--17~d;][]{gonzalez15}, but also longer by a factor of $\gtrsim2$ than what was observed for the luminous Type II-L SN 1998S \citep[$\sim17$~d from explosion;][]{fassia00}. SLSNe I, on the other hand, exhibit rise times similar to the SLSNe II \citep{chen22a}. The two {peculiarly fast-declining SLSNe~II}, SN~2019pud and SN~2020jhm, are also fast to rise (rise times from explosion are $\sim20$ and $\sim17$~d, respectively), and close to the range of non-superluminous SNe II. 

Normal SNe II {(including II-L)} tend to exhibit a plateau or similar phase of flattening in their light curves \citep{anderson14,valenti15}. Only very few SNe II-L show a truly linear light curve after the diffusion peak, such as SN 2016gsd \citep{reynolds20}, which had a peak absolute magnitude of $\sim-20$ mag. Generally speaking, we do not observe the normal sequence of a clear plateau or bump phase, followed by a drop to a radioactive tail phase, in our sample SNe. It is possible that such a phase exists in SN~2019pud and SN~2019uba, but these features could also be{ undulations}, similar to what is often seen in SLSNe~I \citep{hosseinzadeh21,chen22b}. Undulation features in the light curve can be signs of CSI, but these are not observed in 08es-like events or luminous SNe II-L. Our light curves extend to beyond the typical plateau phases of $\lesssim100$~d \citep[see e.g.][]{reynolds20} in six cases; for the remaining five with no clear bumps (as shown in Fig. \ref{fig:george1}), it is possible that the bump or flattening phase occurred after our coverage ended. 

The normal SNe II included in the sample of \citet{anderson14} show an increasing decline rate $s_2$ with brighter peak magnitude. We do not see any such -- or opposite -- trend in our sample, however (see Fig. \ref{fig:peakdelta}): the $s_2$ rates are similar for all SNe studied here regardless of brightness, typically in the range of 1--4~mag~(100 d)$^{-1}$. This is also the range of normal SNe II at peak magnitudes $\lesssim-17.5$ mag. While neither we nor \citet{anderson14} include SNe II between $-19$ and $-20$ mag in our samples, this suggests the $s_2$ trend flattens out after $\sim-18$~mag. We note, though, that $s_2$ normally refers to the plateau/bump phase which our sample generally does not exhibit. 

This lack of a plateau or bump is not the only photometric difference to the luminous SN II-L group {(Fig. \ref{fig:peakdelta})}, despite the spectroscopic similarity: {they} also tend to decline faster at early times than {SLSNe~II}. The plateauless SN 2016gsd \citep{reynolds20}, on the other hand, approaches the decline rates seen in these samples as expressed with the $\Delta g_{50}$ parameter. 
The fast-evolving {SNe~2019pud and 2020jhm} are the exceptions: these events show a decline and a colour evolution in the first 50 days comparable to or even faster than luminous SN~II-L events \citep[especially SN~2013fc;][]{kangas16}, but differ spectroscopically from these events, as opposed to the rest of the sample (see Sect. \ref{sec:spec}). 
All in all, while two luminous SNe II-L, SN 1979C \citep{panagia80} and SN 2013fc \citep{kangas16}, fulfil our criteria as established in Sect. \ref{sec:data}, our sample does not include any SNe that are \emph{both} spectroscopically and photometrically similar to them. 

We note that, in terms of light curves, there is considerable overlap between the ZTF SLSNe IIn, compiled from photometry publicly available through the BTS Sample Explorer\footnote{\url{https://sites.astro.caltech.edu/ztf/bts/explorer.php}} \citep{perley20}, and our sample. This applies especially to the II-P/II-L-like group in Fig. \ref{fig:peakdelta}. Meanwhile, the SLSNe~I show a wide range of photometric properties, also resulting in overlap, but our sample includes relatively fewer fast-declining events. The II-P/II-L-like SNe do not extend to the brightest peak magnitudes, while the bright 08es-like events only overlap with the edge of the SLSN I distribution. Nevertheless, as a whole, SLSNe~II show more similarity to SLSNe~I and especially to SLSNe~IIn in their photometric evolution than to normal SNe~II or luminous SNe~II-L. 

\subsection{Ultraviolet excess}
\label{sec:lc_uv}

\begin{figure*}
\begin{minipage}{\linewidth}
\centering
\includegraphics[width=0.32\columnwidth]{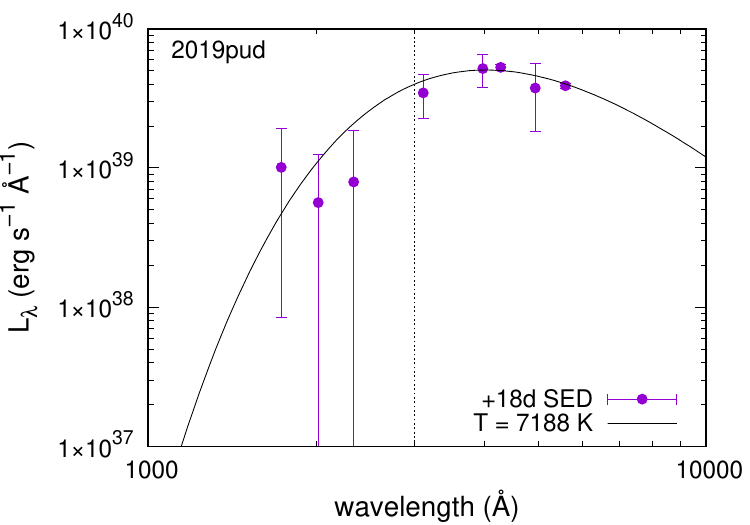}
\includegraphics[width=0.32\columnwidth]{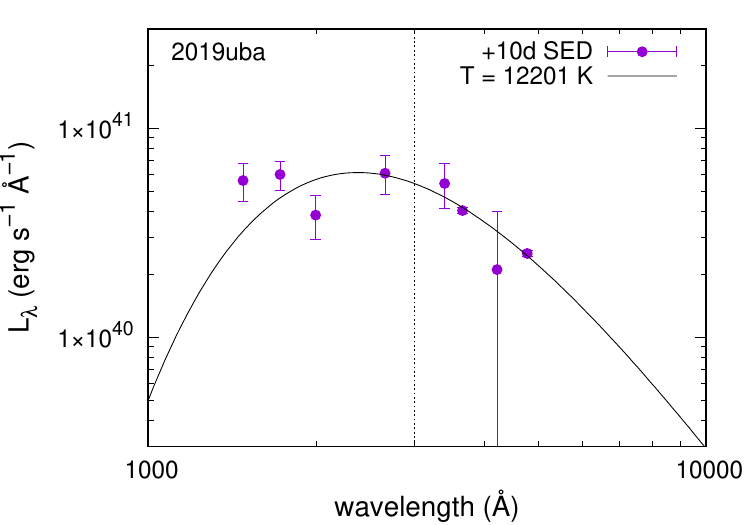} 
\includegraphics[width=0.32\columnwidth]{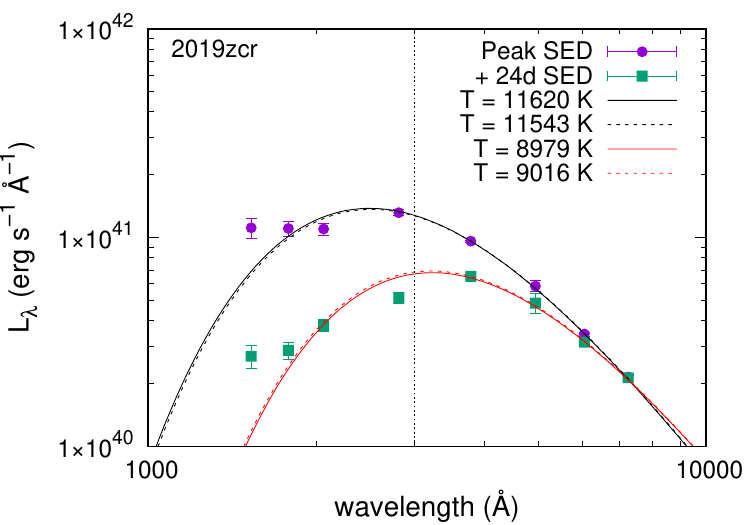} \\
\includegraphics[width=0.32\columnwidth]{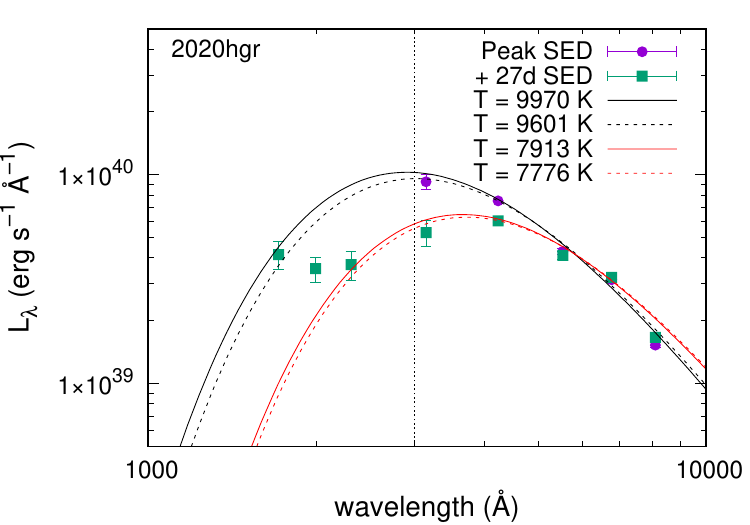}
\includegraphics[width=0.32\columnwidth]{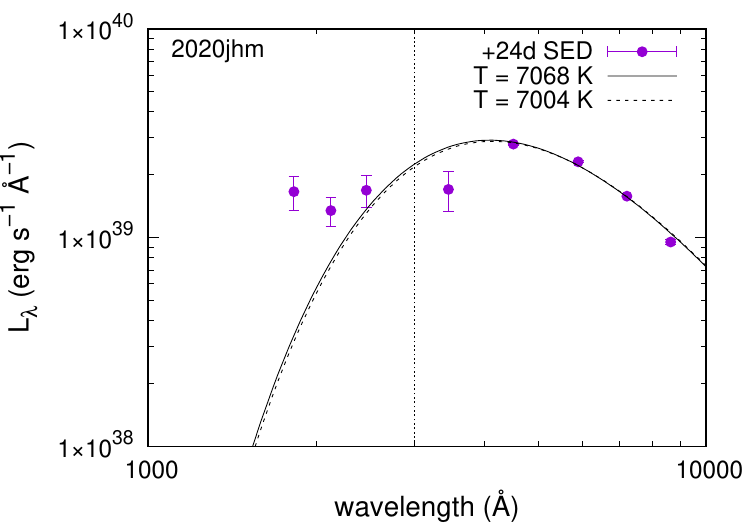}
\includegraphics[width=0.32\columnwidth]{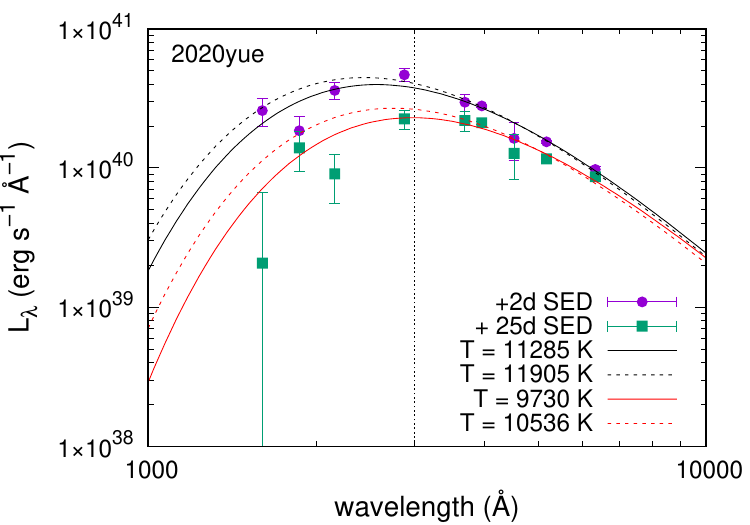} \\
\caption{Blackbody fits (lines) using the combined UV and optical data when available (points). We use the peak epoch if possible; if not, we show the earliest possible epoch. For SNe~2019zcr, 2020hgr and 2020yue we also perform the fit to UV data at $\sim25$ d. The dashed vertical line corresponds to the limit of the modified SED in the {\tt slsn} model at 3000~\AA. Dashed curves correspond to fits without the points below 3000~\AA; these were performed for the SNe where the remaining points have the smallest errors. An excess at the UV wavelengths is immediately clear in four of the six events compared to the blackbody fit.}
\label{fig:BB}
\end{minipage}
\end{figure*}

The spectral energy distributions (SEDs) of SLSNe~I are affected by considerable line blanketing in the ultraviolet (UV) region, which is taken into account in the models that we employ in Sect. \ref{sec:models} \citep[][]{nicholl17}. In order to determine whether this affects our sample as well, we have performed blackbody fits on the six events where we have \emph{Swift} UV photometry available. We show these fits in Fig. \ref{fig:BB}. Clearly, a simple blackbody is not a good description of four of these six SNe; but instead of a UV deficit caused by line blanketing, we see a UV \emph{excess} that, especially in the observed UVW2 band, strengthens with time. In SNe~2019pud and 2020yue the SED is more consistent with a blackbody.

The shape of the UV SED is qualitatively similar to that of the luminous Type II-L SN 1979C \citep{panagia80}. \citet{fransson84} examined this spectrum and considered the UV excess to be a result of CSI; a forest of emission lines from various highly ionized species was caused by ionization and excitation by X-rays from the interaction \citep[a similar UV spectrum was also seen for the Type IIn SN 2010jl and the Type Icn SN~2019hgp:][]{fransson14,dessart15,galyam22}. The events used here show considerable variety in terms of spectra and light curves, including the slowest, fastest, faintest and brightest events in the sample. This suggests that they may represent the bulk of our sample, although this cannot be ascertained. 

\subsection{Radiated energy}
\label{sec:bolo}

We have roughly estimated the total energies radiated in the UV and optical (as we lack infrared data) in our sample SNe through the following steps. Considering the UV excess, we cannot simply fit a blackbody to the observed SEDs. Instead, we have constructed the pseudo-bolometric light curves of SNe~2019zcr and 2020hgr, the objects in the sample with the best UV-to-optical coverage, both showing a clear UV excess. We have used our GP interpolated light curve at each filter to integrate over both wavelength and time using the trapezoidal approximation and setting the flux density to zero at the blue edge of the \emph{UVW2} filter and at the $J$ band. We fit a third-degree polynomial to the bolometric corrections we have obtained for both targets:
\begin{equation}
    L_\mathrm{bol} = L_{gr,\mathrm{RF}} [ A(g-r)^3_\mathrm{RF} + B(g-r)^2_\mathrm{RF} + C(g-r)_\mathrm{RF} + D]~,
\end{equation}
where $L_\mathrm{bol}$ is bolometric luminosity, $L_{gr,\mathrm{RF}}$ is the luminosity in the rest-frame $g$ and $r$ bands and $(g-r)_\mathrm{RF}$ the rest-frame $g-r$ colour. We obtain $A=-42\pm11$, $B=16\pm2$, $C=-2.8\pm0.4$, $D=1.95\pm0.03$.

We have then used these corrections to estimate the UV-to-optical pseudo-bolometric light curves of the less well-observed SNe in our sample and integrate these over time. This assumes all SNe in the sample have similar UV excesses and thus only serves as an order-of-magnitude estimate. The resulting values can be considered rough lower limits for the total radiated energy, as they ignore any infrared and far-UV contribution {and are not} extrapolated in time to unobserved epochs. We list the resulting radiated energies in Table \ref{tab:energy}. 

\begin{table}
    \centering
    \caption{Estimated total energies of the sample SNe radiated in the UV-to-optical range over the \textit{observed} light curve. We assume all SNe here to have a UV excess similar to SNe~2019zcr and 2020hgr, which we use to estimate bolometric corrections.}
    \begin{tabular}{lc}
       SN & Radiated energy \\
       & (erg) \\
        \hline
        SN~2019zcr & $> 3.1\times10^{51}$ \\ 
        SN~2020hgr & $> 3.7\times10^{50}$ \\ 
        \hline
        SN~2018jkq & $\gtrsim 2.3\times10^{50}$ \\
        SN~2019kwr & $\gtrsim 2.4\times10^{50}$ \\ 
        SN~2019cqc & $\gtrsim 1.5\times10^{50}$\\
        SN~2019gsp & $\gtrsim 2.3\times10^{50}$ \\ 
        SN~2019xfs & $\gtrsim 6.0\times10^{50}$ \\ 
        SN~2019pud & $\gtrsim 2.5\times10^{50}$ \\ 
        SN~2018lqi & $\gtrsim 2.2\times10^{50}$ \\ 
        SN~2019aanx & $\gtrsim 9.1\times10^{50}$\\  
        SN~2019uba & $\gtrsim 1.8\times10^{51}$ \\ 
        SN~2020bfe & $\gtrsim 2.6\times10^{50}$ \\ 
        SN~2020jhm & $\gtrsim 1.4\times10^{50}$ \\ 
        SN~2020yue & $\gtrsim 6.1\times10^{50}$ \\
        \hline
    \end{tabular}
    \label{tab:energy}
\end{table} 

\section{Light curve modelling}
\label{sec:models}

\subsection{Modelling setup}
\label{sec:modsetup}

\begin{table}
    \centering
    \caption{Priors of our \texttt{MOSFiT} runs. Each free parameter has either a uniform or log-uniform distribution as indicated.}
    \begin{tabular}{lcc}
       Parameter & Range & Distribution \\
        \hline
        \multicolumn{3}{c}{Common parameters} \\
        \hline
        $n_{H,\mathrm{host}}$ & [$10^{16}$ : $2\times10^{21}$] cm$^{-2}$ & log-uniform \\
        $f_{\mathrm{Ni}}$ & [$10^{-3}$ : 0.3] & log-uniform \\
        $t_{\mathrm{expl}}$ & [-200 : 0] d & uniform \\
        $T_\mathrm{min}$ & [1000 : $10^5$] K & log-uniform \\
        $\kappa$ & $0.34$~cm$^2$~g$^{-1}$ & fixed \\
        $\kappa_\gamma$ & [0.1 : $10^4$]~cm$^2$~g$^{-1}$ & log-uniform \\
        \hline
        \multicolumn{3}{c}{$^{56}$Ni model parameters} \\
        \hline
        $M_\mathrm{ej}$ & [0.1 : 100]~$\mathrm{M}_\odot$ & log-uniform \\
        \hline
        \multicolumn{3}{c}{Magnetar+$^{56}$Ni model parameters} \\
        \hline
        $P_{\mathrm{spin}}$ & [0.7 : 20] ms & uniform \\
        $B_\perp$ & [0.05 : 50]~$\times10^{14}$ G & log-uniform \\
        $M_\mathrm{NS}$ & [1.0 : 2.5]~$\mathrm{M}_\odot$ & uniform \\
        $\theta_\mathrm{PB}$ & [0 : $\pi/2$] rad & uniform \\
        $M_\mathrm{ej}$ & [3 : 100]~$\mathrm{M}_\odot$ & log-uniform \\
        \hline
        \multicolumn{3}{c}{CSI+$^{56}$Ni model parameters} \\
        \hline
        $n$ & 11 & fixed \\
        $\delta$ & 1 & fixed \\
        $s$ & 0 or 2 & fixed \\
        $R_0$ & [0.1 : 1000] AU & log-uniform \\
        $M_\mathrm{CSM}$ & [0.1 : 100]~$\mathrm{M}_\odot$ & log-uniform \\
        $M_\mathrm{ej}$ & [0.1 : 100]~$\mathrm{M}_\odot$ & log-uniform \\
        $\rho$ & [$10^{-15}$ : $10^{-6}$]~cm$^{-3}$ & log-uniform \\
        \hline
    \end{tabular}
    \label{tab:priors}
\end{table}  

\begin{table}
    \centering
    \caption{Ejecta velocity priors of our \texttt{MOSFiT} runs. Each parameter has a uniform distribution between the indicated values.}
    \begin{tabular}{lccc}
       SN & $v_{\mathrm{ej}}$ range \\
       & (km~s$^{-1}$) \\
        \hline
        SN~2018jkq & [6000 : 12000] \\
        SN~2019kwr & [6000 : 12000] \\ 
        SN~2019cqc & [8000 : 16000] \\
        SN~2019gsp & [8000 : 16000] \\ 
        SN~2019xfs & [5000 : 10000] \\ 
        SN~2019pud & [10000 : 20000] \\ 
        SN~2018lqi & [5000 : 10000] \\ 
        SN~2019aanx & [6000 : 12000] \\  
        SN~2019uba & [8000 : 16000] \\ 
        SN~2019zcr & [7000 : 14000] \\ 
        SN~2020bfe & [8000 : 16000] \\ 
        SN~2020hgr & [7000 : 14000] \\ 
        SN~2020jhm &  [12000 : 24000]\\ 
        SN~2020yue & [11000 : 22000] \\
        \hline
    \end{tabular}
    \label{tab:priorvel}
\end{table}  

In order to fit various models to the light curves of our sample SLSNe, we used the publicly available code Modular Open Source Fitter for Transients \citep[\texttt{MOSFiT}\footnote{\url{https://mosfit.readthedocs.io/en/latest/index.html}};][]{guillochon18}. The code includes the $^{56}$Ni decay model \citep{arnett82,nadyozhin94}, labeled \texttt{default}; a combination of CSI and $^{56}$Ni decay based on \citet{chatz13}, labeled \texttt{csmni}; and a combination of a magnetar and $^{56}$Ni decay \citep{nicholl17}, labeled \texttt{magni}. These models are fitted using the dynamic nested sampling package \texttt{dynesty}\footnote{\url{https://dynesty.readthedocs.io/en/latest/}}. We fitted each of these models for each SN. For the {CSI models} we also fixed{ the parameter $s$ (where the CSM density as a function of distance behaves as $\rho \propto r^{-s}$) to two values: $s=0$ (indicating a CSM shell) and $s=2$ (wind-like CSM)}.

We set simple uniform or log-uniform priors for each free parameter. We set the lower limit on the characteristic ejecta velocity to be the FWHM of the H$\alpha$ emission (as this is often measured weeks after the peak), rounded down to the nearest 1000~km~s$^{-1}$, and the upper limit at twice the lower limit, as listed in Table \ref{tab:priorvel}. Near-peak spectra with absorption lines from which to measure photospheric velocities are almost nonexistent {in our sample.} We set the fraction of $^{56}$Ni in the ejecta, $f_{\mathrm{Ni}}$, at a conservative value of $\leq0.3$. Based on the lack of narrow Na~{\sc i}~D absorption lines in the spectra and on the colour variation within the sample described above, we also set an upper limit for the host galaxy extinction, $A_{V,\mathrm{host}} \leq 1$~mag. Host extinction itself is not a parameter in \texttt{MOSFiT}{, but} a related quantity, the column density of neutral hydrogen, $n_{H,\mathrm{host}}$, is. We therefore set the upper limit as $n_{H,\mathrm{host}} \leq 2\times10^{21}$~cm$^{-2}$ based on \citet{guver09}.

The presence of hydrogen in the SN spectra is helpful for setting some of the priors. We fix the Thomson scattering opacity parameter $\kappa = 0.34$~cm$^2$~g$^{-1}$, a typical value for hydrogen-rich ejecta and close to the results of \citet{nagy18}, in each model. In the CSI models, we assume a hydrogen-rich progenitor, but not necessarily an extended envelope such as that of a red supergiant (RSG). Thus the minimum inner radius of the CSM, $R_0$, is set at 0.1 AU ($\sim20 R_\odot$), roughly half the radius of the blue supergiant progenitor of SN 1987A \citep{podsiadlowski92} but larger than a Wolf-Rayet progenitor of a stripped-envelope SN. {In the magnetar models, we can also set a minimum ejecta mass at roughly $3~\mathrm{M}_\odot$. For details on the magnetar model setup, see Appendix \ref{app:model}.}

{Parameters} common to all models include the explosion time before observations $t_\mathrm{expl}$, the opacity to $\gamma$ rays $\kappa_\gamma$ and the minimum temperature $T_\mathrm{min}$. For the magnetar model the free parameters additionally include the spin period $P_\mathrm{spin}$, the magnetic field perpendicular to the spin axis $B_\perp$, the neutron star mass $M_\mathrm{NS}$ and the angle between the magnetic field and spin axis $\theta_\mathrm{PB}$. In the CSI model we additionally include the CSM mass $M_\mathrm{CSM}$ and the CSM density at $R_0$, $\rho$. We fix the density profile parameters in the inner and outer ejecta, $\delta=1$ and $n=11$ respectively. All parameters described above are summarized in Tables \ref{tab:priors} (all parameters except $v_\mathrm{ej}$) and \ref{tab:priorvel} (individual $v_\mathrm{ej}$ for each SN). In total, the \texttt{default} model has 8 free parameters, the \texttt{magni} model has 12 and the \texttt{csmni} model has 11. These numbers include a nuisance parameter $\sigma$, which describes the added variance required to match the model being fitted. We ran each fitting process {until convergence.}

\subsection{Modelling results}

\begin{table*}
\centering    
\begin{minipage}{0.6\linewidth}
    \renewcommand{\arraystretch}{0.83}
    \centering
    \caption{Most important \texttt{MOSFiT} parameters and scores for the $^{56}$Ni powered model. }
    \begin{tabular}{lccccc}
       SN & log $f_{\mathrm{Ni}}$ & log $M_{\mathrm{ej}}$ & log $T_\mathrm{min}$ & $v_{\mathrm{ej}}$ & Score \\
       &  & ($M_{\odot}$) & (K) & (km s$^{-1}$) & \\
        \hline
        SN~2018jkq & $-0.56^{+0.03}_{-0.04}$ & $1.8\pm0.1$ & $4.00\pm0.04$ & $11000^{+700}_{-1300}$ & $-8.5$ \\
        SN~2019kwr & $-0.53\pm0.01$ & $1.38\pm0.02$ & $3.96\pm0.01$ & $11900^{+100}_{-200}$ & 270.7 \\
        SN~2019cqc & $-0.53\pm0.01$ & $1.34\pm0.01$ & $3.87\pm0.01$ & $15900^{+100}_{-200}$ & 398.6 \\
        SN~2019gsp & $-0.57^{+0.03}_{-0.06}$ & $1.8\pm0.2$ & $3.97\pm0.04$ & $14400^{+1200}_{-2000}$ & $-3.3$ \\
        SN~2019xfs & $-0.55^{+0.02}_{-0.04}$ & $1.5\pm0.1$ & $3.89\pm0.02$ & $9400^{+500}_{-900}$ & $-37.3$ \\
        SN~2019pud & $-0.6\pm0.1$ & $1.8\pm0.2$ & $3.84\pm0.03$ & $16700^{+2100}_{-3100}$ & $-15.5$ \\
        SN~2018lqi & $-0.56^{+0.02}_{-0.04}$ & $1.84\pm0.09$ & $3.98\pm0.04$ & $9200^{+600}_{-900}$ & $-0.1$\\
        SN~2019aanx & $-0.53\pm0.01$ & $1.99\pm0.01$ & $3.96\pm0.02$ & $11800^{+200}_{-300}$ & $57.9$\\
        SN~2019uba & $-0.53\pm0.01$ & $1.99\pm0.01$ & $4.04\pm0.02$ & $15700^{+200}_{-500}$ & $-33.7$\\
        SN~2019zcr & $-0.53\pm0.01$ & $1.99\pm0.01$ & $4.02\pm0.02$ & $13800^{+200}_{-300}$ & $-180.3$\\
        SN~2020bfe & $-0.53\pm0.01$ & $1.35\pm0.02$ & $3.86\pm0.01$ & $15600^{+300}_{-500}$ & 147.7\\
        SN~2020hgr & $-0.53\pm0.01$ & $1.38\pm0.01$ & $3.98\pm0.01$ & $13900^{+100}_{-200}$ & 332.9\\
        SN~2020jhm & $-0.52\pm0.01$ & $0.64\pm0.01$ & $3.89\pm0.01$ & $23900\pm100$ & $-34.6$ \\
        SN~2020yue & $-0.52\pm0.01$ & $1.99\pm0.01$ & $4.02\pm0.01$ & $23900^{+100}_{-200}$ & 187.7\\
        \hline
    \end{tabular}
    \label{tab:resultsNi}
\end{minipage}
\end{table*}

\begin{table*}
\centering    
\begin{minipage}{0.95\linewidth}
    \renewcommand{\arraystretch}{0.83}
    \centering
    \caption{Most important \texttt{MOSFiT} parameters and scores for the magnetar-powered model with $^{56}$Ni. }
    \begin{tabular}{lcccccccc}
       SN & log $B_\perp$ & $M_{\mathrm{NS}}$ & $P_{\mathrm{spin}}$ & log $f_{\mathrm{Ni}}$ & log $M_{\mathrm{ej}}$ & $\theta_{PB}$ & $v_{\mathrm{ej}}$ & Score\\
       & ($10^{14}$ G) & ($M_{\odot}$) & (ms) &  & ($M_{\odot}$) & (rad) & (km s$^{-1}$) & \\
        \hline
        SN~2018jkq & $0.5^{+0.2}_{-0.3}$ & $1.7\pm0.4$ & $1.9^{+0.9}_{-0.7}$ & $-2.1^{+0.7}_{-0.6}$ & $0.9\pm0.1$ & $0.8^{+0.4}_{-0.3}$ & $7700\pm500$ & 53.7 \\
        SN~2019kwr & $0.4\pm0.2$ & $1.3^{+0.3}_{-0.2}$ & $1.5^{+0.9}_{-0.5}$ & $-2.4\pm0.4$ & $0.96^{+0.04}_{-0.06}$ & $0.9^{+0.4}_{-0.3}$ & $7500^{+200}_{-300}$ & 741.5 \\
        SN~2019cqc & $0.9\pm0.2$ & $2.1^{+0.3}_{-0.4}$ & $2.0^{+0.8}_{-0.7}$ & $-0.54^{+0.01}_{-0.02}$ & $0.93\pm0.01$ & $0.7\pm0.2$ & $8100\pm100$ & 724.7\\
        SN~2019gsp & $0.7\pm0.2$ & $1.9\pm0.4$ & $2.0^{+1.1}_{-0.9}$ & $-2.1^{+0.5}_{-0.6}$ & $0.65\pm0.05$ & $1.0\pm0.3$ & $8500^{+600}_{-400}$ & 66.8\\
        SN~2019xfs & $0.2\pm0.2$ & $1.9^{+0.4}_{-0.4}$ & $1.0^{+0.3}_{-0.2}$ & $-1.6^{+0.4}_{-0.5}$ & $0.96\pm0.04$ & $0.9\pm0.3$ & $5800^{+200}_{-300}$ & 58.4\\
        SN~2019pud & $0.8\pm0.3$ & $1.7\pm0.5$ & $3\pm2$ & $-0.6^{+0.1}_{-0.2}$ & $0.53^{+0.05}_{-0.04}$ & $0.9\pm0.4$ & $16700^{+1900}_{-1500}$ & 13.1\\
        SN~2018lqi & $0.2\pm0.3$ & $1.8\pm0.5$ & $5.0^{+1.0}_{-1.1}$ & $-1.4^{+0.6}_{-1.0}$ & $0.6\pm0.1$ & $0.9\pm0.4$ & $6200\pm400$ & 72.2\\
        SN~2019aanx & $-0.2\pm0.3$ & $2.0^{+0.4}_{-0.5}$ & $2.6^{+0.5}_{-0.6}$ & $-1.8^{+0.9}_{-0.8}$ & $1.0^{+0.3}_{-0.2}$ & $0.9^{+0.5}_{-0.5}$ & $10200\pm700$ & 117.8\\
        SN~2019uba & $-0.1^{+0.2}_{-0.3}$ & $1.7\pm0.4$ & $2.5\pm0.6$ & $-2.2^{+0.8}_{-0.5}$ & $0.51^{+0.04}_{-0.02}$ & $0.9\pm0.3$ & $9300^{+1200}_{-700}$ & 65.7\\
        SN~2019zcr & $-0.2\pm0.2$ & $1.8^{+0.5}_{-0.6}$ & $2.0^{+0.4}_{-0.5}$ & $-2.1^{+0.8}_{-0.6}$ & $0.69\pm0.03$ & $0.6\pm0.4$ & $11300\pm300$ & 232.7\\
        SN~2020bfe & $-0.0^{+0.2}_{-0.3}$ & $1.8\pm0.5$ & $5.7^{+1.0}_{-1.1}$ & $-1.4^{+0.6}_{-1.1}$ & $0.89\pm0.03$ & $1.1^{+0.3}_{-0.4}$ & $8100^{+200}_{-100}$ & 215.6\\
        SN~2020hgr & $-0.1\pm0.2$ & $2.2^{+0.3}_{-0.4}$ & $4.2^{+1.4}_{-0.9}$ & $-0.7^{+0.1}_{-0.2}$ & $1.3\pm0.1$ & $0.9\pm0.4$ & $12200^{+1400}_{-2400}$ & 354.6\\
        SN~2020jhm & $1.1^{+0.1}_{-0.2}$ & $2.1^{+0.3}_{-0.5}$ & $0.8^{+0.3}_{-0.1}$ & $-1.02\pm0.05$ & $0.54\pm0.02$ & $1.2^{+0.2}_{-0.3}$ & $16900^{+600}_{-800}$ & 780.3\\
        SN~2020yue & $-0.3\pm0.2$ & $1.7^{+0.4}_{-0.5}$ & $3.5^{+0.6}_{-0.7}$ & $-1.6^{+0.7}_{-0.8}$ & $0.8^{+0.1}_{-0.2}$ & $1.0\pm0.3$ & $18000^{+3700}_{-4600}$ & 332.2\\
        \hline
    \end{tabular}
    \label{tab:resultsMagNi}
\end{minipage}
\end{table*}

\begin{table*}
\centering    
\begin{minipage}{0.95\linewidth}
    \renewcommand{\arraystretch}{0.83}
    \centering
    \caption{Most important \texttt{MOSFiT} parameters and scores for the CSM+Ni model when $s=2$. The kinetic energy $E_k$ is calculated separately as $E_k = 0.3M_{\mathrm{ej}}v_{\mathrm{ej}}^2$ and is not a model parameter.}
    \begin{tabular}{lcccccccc}
       SN & log $f_{\mathrm{Ni}}$ & log $M_{\mathrm{CSM}}$ & log $M_{\mathrm{ej}}$ & log $R_0$ & log $\rho$ & $v_{\mathrm{ej}}$ & $E_k$ & Score \\
       & & ($M_{\odot}$) & ($M_{\odot}$) & (AU) & (g cm$^{-3}$) & (km s$^{-1}$) & ($10^{51}$~erg) & \\
        \hline
        SN~2018jkq & $-0.8^{+0.2}_{-0.3}$ & $0.6^{+0.3}_{-0.2}$ & $0.6\pm0.5$ & $1.1\pm0.6$ & $-10.7^{+1.0}_{-0.9}$ & $8000\pm500$ & $1.7^{+3.0}_{-1.1}$ & 62.1 \\
        SN~2019kwr & $-1.7\pm0.2$ & $-0.38^{+0.06}_{-0.05}$ & $1.62\pm0.04$ & $0.84^{+0.06}_{-0.07}$ & $-10.1\pm0.2$ & $7700\pm200$ & $15\pm2$ &  739.2 \\
        SN~2019cqc & $-1.3^{+0.7}_{-1.4}$ & $-0.1^{+0.3}_{-0.1}$ & $0.3^{+0.3}_{-0.5}$ & $1.4^{+0.3}_{-0.1}$ & $-9.1^{+0.1}_{-0.2}$ & $8200\pm200$ & $0.8\pm0.5$ &  696.4\\
        SN~2019gsp & $-1.8^{+0.9}_{-0.7}$ & $-0.2^{+0.3}_{-0.4}$ & $0.4^{+0.6}_{-0.7}$ & $1.6^{+0.3}_{-0.4}$ & $-9.3^{+1.1}_{-0.40}$ & $8600^{+500}_{-400}$ & $1.0^{+3.1}_{-0.9}$ &  72.2 \\
        SN~2019xfs & $-1.6^{+0.7}_{-0.9}$ & $0.5\pm0.4$ & $0.1^{+0.6}_{-0.7}$ & $2.0\pm0.4$ & $-8.4^{+0.5}_{-0.4}$ & $5300^{+300}_{-200}$ & $3.2^{+1.1}_{-0.8}$ &  52.4 \\
        SN~2019pud & $-1.1^{+0.2}_{-0.1}$ & $-0.7^{+0.3}_{-0.2}$ & $1.4\pm0.2$ & $1.3^{+0.3}_{-0.2}$ & $-10.1^{+1.2}_{-1.1}$ & $12700\pm900$ & $23^{+10}_{-8}$ & 15.2\\
        SN~2018lqi & $-1.8\pm0.8$ & $-0.3\pm0.4$ & $-0.0\pm0.7$ & $1.3\pm0.4$ & $-8.1^{+0.9}_{-0.5}$ & $6400^{+500}_{-400}$ & $0.2^{+0.9}_{-0.2}$ & 68.4\\
        SN~2019aanx & $-2.0^{+1.0}_{-0.7}$ & $1.29^{+0.10}_{-0.04}$ & $-0.1^{+0.4}_{-0.2}$ & $-0.2^{+2.2}_{-0.6}$ & $-8^{+2}_{-5}$ & $11200^{+600}_{-800}$ & $0.6^{+0.5}_{-0.2}$ &  120.0\\
        SN~2019uba & $-0.56^{+0.02}_{-0.04}$ & $0.42^{+0.06}_{-0.07}$ & $1.9^{+0.1}_{-0.2}$ & $0.2^{+0.8}_{-0.7}$ & $-9\pm2$ & $11100\pm400$ & $53\pm4$ & 88.3 \\
        SN~2019zcr & $-1.8^{+0.6}_{-0.8}$ & $1.4\pm0.2$ & $0.1\pm0.5$ & $2.6^{+0.3}_{-0.4}$ & $-13.3^{+0.5}_{-0.3}$ & $11900\pm200$ & $1.0^{+1.7}_{-0.7}$ & 228.1\\
        SN~2020bfe & $-1.1^{+0.2}_{-0.4}$ & $0.6^{+0.1}_{-0.2}$ & $0.4^{+0.3}_{-0.2}$ & $1.7^{+0.1}_{-0.2}$ & $-11.0^{+0.4}_{-0.3}$ & $8400\pm200$ & $1.1^{+0.8}_{-0.3}$ &  181.4\\
        SN~2020hgr & $-0.7^{+0.2}_{-0.9}$ & $-0.4^{+0.5}_{-0.4}$ & $1.1^{+0.2}_{-0.9}$ & $0.8^{+0.5}_{-0.4}$ & $-7.7\pm0.4$ & $7700^{+800}_{-500}$ & $3.9^{+1.9}_{-3.4}$ & 310.1\\
        SN~2020jhm & $-0.9^{+0.2}_{-0.9}$ & $-0.3^{+0.1}_{-0.5}$ & $0.2^{+0.9}_{-0.2}$ & $1.7^{+0.1}_{-0.5}$ & $-11.0^{+0.9}_{-0.4}$ & $12500^{+300}_{-200}$ & $1.6^{+8.5}_{-0.5}$ &  854.9\\
        SN~2020yue & $-1.9^{+0.7}_{-0.6}$ & $-0.3^{+0.6}_{-0.3}$ & $0.9^{+0.5}_{-0.9}$ & $1.1^{+0.5}_{-0.3}$ & $-9.4^{+0.3}_{-0.7}$ & $13400^{+1200}_{-1100}$ & $9^{+14}_{-8}$ &  345.4\\
        \hline
    \end{tabular}
    \label{tab:resultsCSMNi2}
\end{minipage}
\end{table*}

\begin{table*}
\centering    
\begin{minipage}{0.95\linewidth}
    \renewcommand{\arraystretch}{0.83}
    \centering
    \caption{Most important \texttt{MOSFiT} parameters and scores for the CSM+Ni model when $s=0$. The kinetic energy $E_k$ is calculated separately as $E_k = 0.3M_{\mathrm{ej}}v_{\mathrm{ej}}^2$ and is not a model parameter.}
    \begin{tabular}{lcccccccc}
       SN & log $f_{\mathrm{Ni}}$ & log $M_{\mathrm{CSM}}$ & log $M_{\mathrm{ej}}$ & log $R_0$ & log $\rho$ & $v_{\mathrm{ej}}$ & $E_k$ & Score \\
       & & ($M_{\odot}$) & ($M_{\odot}$) & (AU) & (g cm$^{-3}$) & (km s$^{-1}$) & ($10^{51}$~erg) & \\
        \hline
        SN~2018jkq & $-1.2\pm0.4$ & $0.7\pm0.2$ & $0.8^{+0.3}_{-0.2}$ & $1.0\pm0.8$ & $-11.7^{+0.6}_{-0.4}$ & $8800^{+600}_{-1100}$ & $2.7^{+1.6}_{-1.1}$ &  53.3\\
        SN~2019kwr & $-2.2^{+0.3}_{-0.6}$ & $0.9\pm0.2$ & $1.16^{+0.06}_{-0.05}$ & $2.4\pm0.2$ & $-10.96^{+0.06}_{-0.07}$ & $6800\pm200$ & $4.0^{+0.6}_{-0.5}$ &  730.9\\
        SN~2019cqc & $-1.6^{+0.7}_{-0.9}$ & $1.2^{+0.2}_{-0.3}$ & $-0.3^{+0.6}_{-0.5}$ & $2.5^{+0.2}_{-0.3}$ & $-6.7^{+0.4}_{-0.5}$ & $8000\pm100$ & $0.2^{+0.5}_{-0.2}$ & 694.7\\
        SN~2019gsp & $-2.8\pm0.2$ & $-0.2^{+0.2}_{-0.4}$ & $0.9\pm0.2$ & $0.4^{+0.4}_{-0.6}$ & $-11.5^{+0.8}_{-0.4}$ & $9900\pm400$ & $4.3\pm1.0$ &  67.7\\
        SN~2019xfs & $-2.5\pm0.4$ & $0.81^{+0.06}_{-0.04}$ & $1.60\pm0.06$ & $1.3^{+0.9}_{-1.4}$ & $-12.7^{+0.2}_{-0.1}$ & $5900^{+200}_{-300}$ & $9^{+2}_{-1}$ & 52.6\\
        SN~2019pud & $-0.70\pm0.07$ & $-0.34\pm0.08$ & $1.0\pm0.2$ & $0.4^{+0.7}_{-0.8}$ & $-12.5\pm0.2$ & $11000^{+700}_{-600}$ & $8\pm3$ & 18.4\\
        SN~2018lqi & $-1.4^{+0.5}_{-0.6}$ & $0.76^{+0.04}_{-0.06}$ & $1.19^{+0.05}_{-0.06}$ & $0.3^{+1.1}_{-0.9}$ & $-12.61^{+0.06}_{-0.09}$ & $5500\pm200$ & $2.8\pm0.3$ &  79.7 \\
        SN~2019aanx & $-1.9\pm0.8$ & $1.37\pm0.04$ & $0.8\pm0.2$ & $1.0^{+1.0}_{-1.1}$ & $-12.9\pm0.1$ & $9500\pm700$ & $3.4^{+1.1}_{-0.9}$ & 122.2\\
        SN~2019uba & $-0.55^{+0.02}_{-0.03}$ & $-0.1\pm0.1$ & $1.64\pm0.07$ & $0.4^{+0.6}_{-0.7}$ & $-11.7\pm0.2$ & $9400^{+800}_{-700}$ & $23\pm2$ & 85.2\\
        SN~2019zcr & $-1.1\pm0.4$ & $1.2^{+0.2}_{-0.1}$ & $1.4^{+0.1}_{-0.2}$ & $2.5\pm0.2$ & $-13.1^{+0.3}_{-0.1}$ & $10200\pm300$ & $13^{+2}_{-5}$ & 236.9\\
        SN~2020bfe & $-1.8\pm0.8$ & $1.3\pm0.3$ & $-0.1\pm0.6$ & $2.4^{+0.2}_{-0.3}$ & $-7.4^{+0.5}_{-0.6}$ & $8100^{+200}_{-100}$ & $0.3^{+0.9}_{-0.3}$ & 200.4 \\
        SN~2020hgr & $-1.7\pm0.8$ & $1.49\pm0.04$ & $1.1^{+0.2}_{-0.4}$ & $0.9\pm1.2$ & $-13.0\pm0.1$ & $8100^{+900}_{-600}$ & $5^{+2}_{-3}$ &  380.3 \\
        SN~2020jhm & $-1.38^{+0.03}_{-0.02}$ & $-0.93^{+0.02}_{-0.03}$ & $0.66\pm0.02$ & $-0.3^{+0.6}_{-0.5}$ & $-10.82^{+0.06}_{-0.03}$ & $12100^{+200}_{-100}$ & $4.0\pm0.2$ & 853.6\\
        SN~2020yue & $-1.6^{+0.7}_{-0.8}$ & $1.0\pm0.3$ & $0.2^{+0.6}_{-0.8}$ & $2.2\pm0.3$ & $-9.2^{+0.7}_{-0.6}$ & $12200^{+200}_{-100}$ & $1.4^{+3.3}_{-1.2}$ &  325.8\\
        \hline
    \end{tabular}
    \label{tab:resultsCSMNi0}
\end{minipage}
\end{table*}

We include the light curves and corner plots for our \texttt{MOSFiT} modelling as supplementary material, available online. The median posterior parameter values and their $1\sigma$ errors for each SN and model are presented in Tables \ref{tab:resultsNi} -- \ref{tab:resultsCSMNi0}. \texttt{MOSFiT} determines a likelihood score \citep[based on the Watanabe-Akaike Information Criterion or WAIC;][]{watanabe10} for each posterior ensemble. Higher values indicate a better fit, and scores are comparable between models with different numbers of free parameters \citep[see][]{guillochon18}. 

{$^{56}$Ni decay alone cannot reproduce our light curves.} The nickel fraction $f_{\mathrm{Ni}}$ gravitates toward its maximum allowed value. Combined with the large required ejecta masses (typically tens of $\mathrm{M}_\odot$), this results in extremely high $^{56}$Ni masses{, while} the observed light curve evolution is faster than such large ejecta masses would require. The likelihood scores determined by \texttt{MOSFiT} are the lowest for the $^{56}$Ni model, with the exception of SN~2020hgr, for which we do not obtain a good fit with any model in the UV (see below). Therefore, we rule out $^{56}$Ni decay as the dominant power source. 

A few objects with UV data are problematic for all of the models; these are SNe~2020hgr, 2020jhm and 2020yue. SNe~2020hgr and 2020yue show a fast-declining UV light curve that is not reproduced by either the magnetar or CSM models, while UV points of SN~2020jhm are under-predicted in both models. This does not apply to all UV data, however: SNe~2019xfs, 2019pud, 2019uba and 2019zcr also include UV data and are reproduced by both magnetar and CSM models. This is despite the observed UV excess in SN~2019zcr. The UV discrepancy does not necessarily pose a problem: both models could have trouble reproducing the UV light curve if, e.g., the one-dimensional \citet{chatz13} model cannot account for all the mechanisms at play in the interaction. It has been shown that this model can produce light curves an order of magnitude different than those from more detailed, numerical CSI models \citep{sorokina16}. Overall, most of the sample remains consistent with both magnetar and CSI {power sources}. The likelihood scores are lowest for $^{56}$Ni decay, but the difference between the CSI and magnetar scores is $\lesssim20$~\%, usually $<10$~\%. Therefore we cannot distinguish between these power sources based on the score, nor between different CSI models where $s=0$ or $s=2$. A similar ambiguity in the power source based on light curves alone was found by \citet{chen22b} for a large sample of SLSNe~I. 

\section{Host galaxy properties}
\label{sec:hosts}

The properties of the host galaxies of the SNe in our sample can shed light on their progenitors. The hosts of previously studied SLSNe II were faint, presumably metal-poor dwarf galaxies \citep{inserra18,Schulze2018a} similar to those of SLSNe I, whereas SLSNe IIn occupy a wider range in metallicity, stellar mass and brightness \citep{perley16,Schulze2021a}. Here we perform a comparison between previous studies of SN host galaxies and our sample by fitting stellar population models to host galaxy photometry.

We have retrieved science-ready coadded images of the host galaxies from the \textit{Galaxy Evolution Explorer} (\emph{GALEX}) general release 6/7 \citep{Martin2005a}, the Sloan Digital Sky Survey data release 9 \citep[SDSS DR 9;][]{Ahn2012a}, DESI Legacy Imaging Surveys \citep[Legacy Surveys, LS;][]{Dey2018a} data release 8, the data archive of the 3.6~m Canada-France-Hawaii Telescope (USA), and \emph{WISE} images \citep{Wright2010a} from the unWISE archive \citep{Lang2014a}\footnote{\url{http://unwise.me}}. The unWISE images are based on the public \emph{WISE} data and include images from the ongoing NEOWISE-Reactivation mission R3 \citep{Mainzer2014a, Meisner2017a}. For SNe~2019cqc, 2020jhm and 2020yue, we augmented the SEDs with UV and optical data obtained with the \emph{Swift}/UVOT in October 2021, after the SNe had faded. 

The brightness of each host galaxy was measured using \texttt{LAMBDAR}\footnote{\url{https://github.com/AngusWright/LAMBDAR}} \citep[Lambda Adaptive Multi-Band Deblending Algorithm in R;][]{Wright2016a} and the methods described in \citet{Schulze2021a}. The photometry on the UVOT images was done with \texttt{uvotsource} in \texttt{HEASoft} and using an aperture encircling the entire galaxy. All magnitudes were transformed into the AB system using \citet{Breeveld2011a} and {\citet[][their Table 3]{Cutri2013a}}. In the case of SN~2019xfs, which is located $\sim1\farcs8$ from an 18th magnitude star, we removed the star {with \texttt{galfit} \citep{Peng2013a}. We} measured the flux at the explosion site using the aperture photometry tool presented in \citet{Schulze2018a}. 

For the fitting itself, we used the \texttt{Prospector} package\footnote{\url{https://github.com/bd-j/prospector}}, version 0.3 \citep{Leja2017a}, to model the SEDs of the host galaxies and extract their physical parameters. \texttt{Prospector} uses the Flexible Stellar Population Synthesis (\texttt{FSPS}\footnote{\url{https://github.com/cconroy20/fsps}}) code \citep{Conroy2009a} for the physical model and \texttt{python-fsps}\footnote{\url{https://dfm.io/python-fsps/current/}} \citep{ForemanMackey2014a} for a Python-based interface. For details about the model setup, see \citet{Schulze2021a}; we performed these fits in an identical manner with the same assumptions and priors.

\begin{table*}
\centering    
\begin{minipage}{0.95\linewidth}
    \renewcommand{\arraystretch}{0.83}
    \caption{Results from the host galaxy SED modelling with \texttt{Prospector}. The absolute magnitudes are not corrected for host reddening. The SFRs are corrected for host reddening. For the host attenuation $E_{\rm host} (B-V)$ we used the \citet{Calzetti2000a} model. The age refers to the age of the stellar population. }
    \begin{tabular}{lcccccccc}
    \centering
       SN	& Redshift	& $\chi^2/{\rm d.o.f.}$	& $E_{\rm host}(B-V)$	& $ M_{B}$	&$\log~{\rm SFR}$			&$\log~M$		&$\log~{\rm sSFR}$	& Age\\ 
			& 			& 							& (mag)					& (mag)	 		&$(\mathrm{M}_\odot\,{\rm yr}^{-1})$	&($\mathrm{M}_\odot$)	&$({\rm yr}^{-1})$		&$({\rm Gyr})$ \\ 
        \hline
        SN~2018jkq & 0.119 & $25.56/20$ & $0.10^{+0.08}_{-0.05}$ & $-20.73^{+0.08}_{-0.04}$ & $0.0^{+0.3}_{-0.2}$ & $10.2^{+0.2}_{-0.1}$ & $-10.2^{+0.4}_{-0.2}$ & $2.4^{+2.4}_{-1.0}$ \\ 
        SN~2019kwr & 0.202 & $9.77/9$ & $0.3^{+0.2}_{-0.1}$ & $-18.8^{+0.2}_{-0.1}$ & $0.2^{+0.5}_{-0.3}$ & $9.5^{+0.2}_{-0.4}$ & $-9.3^{+0.9}_{-0.4}$ & $4.4^{+5.4}_{-3.8}$ \\ 
        SN~2019cqc & 0.117 & $28.05/15$ & $0.12^{+0.06}_{-0.05}$ & $-19.85^{+0.08}_{-0.04}$ & $0.1\pm0.2$ & $9.8^{+0.2}_{-0.3}$ & $-9.7^{+0.6}_{-0.2}$ & $7^{+5}_{-6}$ \\ 
        SN~2019gsp & 0.171 & $0.73/3$ & $0.3^{+0.3}_{-0.2}$ & $-16.2^{+0.3}_{-0.2}$ & $-0.5^{+0.8}_{-0.6}$ & $7.9^{+0.5}_{-0.7}$ & $-8.4^{+1.2}_{-0.9}$ & $0.5^{+2.8}_{-0.5}$ \\ 
        SN~2019xfs & 0.116 & $0.70/2$ & $0.6^{+0.4}_{-0.3}$ & $-16.5^{+0.7}_{-0.4}$ & $0.3^{+1.0}_{-0.9}$ & $7.9^{+0.7}_{-0.6}$ & $-7.5^{+1.2}_{-1.4}$ & $0.1^{+1.4}_{-0.1}$ \\ 
        SN~2019pud & 0.114 & $3.85/2$ & $0.9^{+0.8}_{-0.7}$ & $-12^{+5}_{-2}$ & $-1.1^{+1.3}_{-1.5}$ & $6.9^{+1.4}_{-1.3}$ & $-8\pm2$ & $0.3^{+4.0}_{-0.3}$ \\ 
        SN~2018lqi & 0.202 & $9.55/7$ & $0.2^{+0.2}_{-0.1}$ & $-18.57^{+0.09}_{-0.05}$ & $-0.1^{+0.5}_{-0.3}$ & $9.1^{+0.2}_{-0.3}$ & $-9.2^{+0.8}_{-0.4}$ & $4^{+5}_{-3}$ \\ 
        SN~2019aanx & 0.403 & $1.44/3$ & $0.2^{+0.4}_{-0.2}$ & $-18.8^{+0.4}_{-0.3}$ & $0.6^{+0.8}_{-0.5}$ & $8.3^{+0.8}_{-0.7}$ & $-7.6^{+1.1}_{-1.0}$ & $0.1^{+0.8}_{-0.1}$ \\ 
        SN~2019uba & 0.303 & $2.51/3$ & $0.6^{+0.3}_{-0.2}$ & $-17.3^{+0.5}_{-0.3}$ & $0.7^{+0.5}_{-0.7}$ & $8.6^{+0.8}_{-0.9}$ & $-7.8^{+1.1}_{-1.2}$ & $0.1^{+1.6}_{-0.1}$ \\ 
        SN~2019zcr & 0.26 & $9.94/3$ & $0.8^{+0.6}_{-0.50}$ & $-15.5^{+0.8}_{-0.5}$ & $0.3^{+1.3}_{-1.2}$ & $8.6^{+0.9}_{-1.2}$ & $-8.2^{+1.4}_{-1.3}$ & $0.4^{+3.5}_{-0.4}$ \\ 
        SN~2020bfe & 0.099 & $9.42/12$ & $0.17^{+0.08}_{-0.04}$ & $-20.3^{+0.3}_{-0.1}$ & $0.4\pm0.2$ & $10.1^{+0.2}_{-0.4}$ & $-9.7^{+0.7}_{-0.2}$ & $7\pm5$ \\ 
        SN~2020hgr & 0.126 & $2.98/8$ & $0.1^{+0.2}_{-0.1}$ & $-17.1^{+0.2}_{-0.1}$ & $-0.6^{+0.4}_{-0.3}$ & $8.0^{+0.4}_{-0.5}$ & $-8.5^{+0.7}_{-0.6}$ & $0.7^{+1.8}_{-0.6}$ \\ 
        SN~2020jhm & 0.057 & $23.74/16$ & $0.13^{+0.08}_{-0.04}$ & $-19.2^{+0.4}_{-0.1}$ & $-0.3^{+0.4}_{-0.2}$ & $9.6^{+0.2}_{-0.5}$ & $-9.9^{+1.0}_{-0.3}$ & $4^{+5}_{-4}$ \\ 
        SN~2020yue & 0.204 & $15.86/20$ & $0.07^{+0.14}_{-0.06}$ & $-20.29\pm0.03$ & $0.2^{+0.4}_{-0.2}$ & $10.4^{+0.1}_{-0.3}$ & $-10.1^{+0.6}_{-0.3}$ & $11^{+3}_{-6}$ \\ 
        \hline
    \end{tabular}
    \label{tab:host_properties}
\end{minipage}
\end{table*}

\begin{table}
    \caption{Results ($p$-values) of our Anderson-Darling tests between the listed host galaxy properties of our sample SNe and earlier published SLSNe II from \citet{Schulze2018a} (total N=17), and comparison subsamples of \citet{Schulze2021a}.}
    \begin{tabular}{lccccc}
    \centering
        & & \multicolumn{4}{c}{SLSN II host properties} \\
       Host sample & N & $M_B$ & log $M$ & log SFR & log sSFR \\
        \hline
        SLSNe IIn & 14 & 0.85 & 0.86 & 0.63 & 0.70 \\
        SLSNe I & 36 & 0.36 & 0.17 & 0.68 & 0.03 \\
        SNe II ($z>0.08$) & 51 & 0.17 & 0.13 & 0.46 & 0.02 \\
        SNe IIn ($z>0.08$) & 48 & 0.25 & 0.14 & 0.97 & 0.05 \\
        SNe Ibc ($z>0.08$) & 31 & 0.20 & 0.52 & 0.84 & 0.17 \\
        \hline
    \end{tabular}
    \label{tab:adtest}
\end{table}  

\begin{figure}
\centering
\includegraphics[width=0.99\columnwidth]{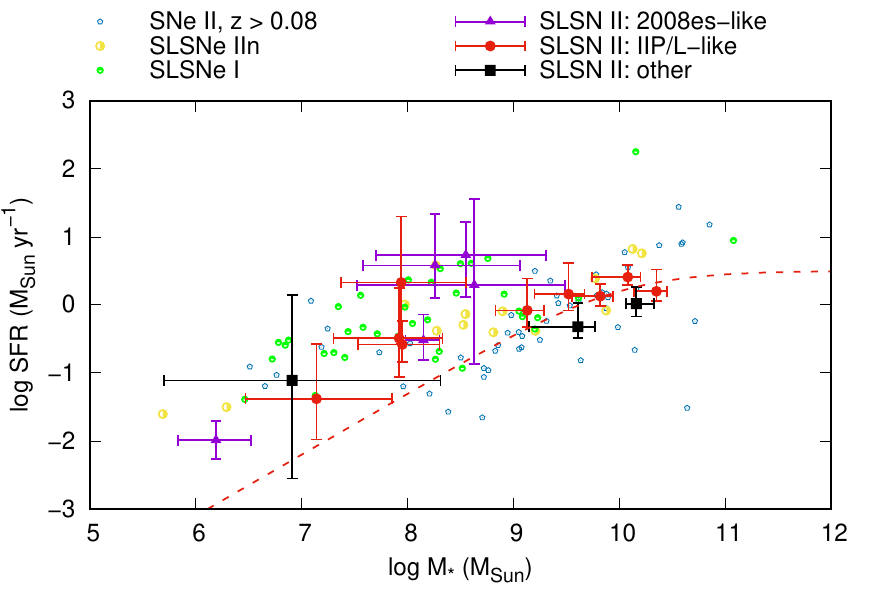}
\caption{Host galaxies of SLSNe II compared to those of SLSNe I and IIn and normal SNe II from \citet{Schulze2021a} in terms of stellar mass and SFR. We also include the hosts of SNe 2008es and 2013hx \citep{Schulze2018a}. The red dashed line corresponds to the galaxy main sequence \citep[with parameters extrapolated to $z=0.2$;][]{lee15}.}
\label{fig:hosts}
\end{figure}

The results of the fitting process are listed for each host galaxy in Table \ref{tab:host_properties}. {We compare} the stellar masses and SFRs of our sample to those of SLSNe I, SLSNe IIn and normal SNe II from \citet{Schulze2021a} in Fig. \ref{fig:hosts}, also including the host galaxies of SNe 2008es and 2013hx from \citet{Schulze2018a}. To address cosmic evolution and make the normal SN II sample more comparable to ours, we only include the SNe II with $z>0.08$, corresponding roughly to the most distant 10\%. The SLSN II hosts in general overlap strongly with those of both SLSNe IIn and SLSNe I. While there is overlap with normal SNe II as well, the SLSN II hosts preferentially seem to be somewhat less massive and more strongly star-forming than them or the galaxy main sequence \citep{lee15}.

We have investigated possible differences using two-sample Anderson-Darling tests with the distributions of absolute magnitude, stellar mass, SFR and specific SFR (sSFR); here we have also included the host galaxies of normal SNe Ibc and IIn from \citet{Schulze2021a}, applying the same distance cut of $z>0.08$. The test results are listed in Table \ref{tab:adtest}. The host galaxies of SLSNe II are consistent with those of SLSNe IIn in terms of all four properties, and the $p$-values are all $>0.6$, suggesting a strong overlap. The sSFRs of the SLSN II hosts are {individually} inconsistent at the 95\% confidence level with being drawn from the same distributions as the sSFRs of SLSN I and SN II hosts{. However, the results are affected by the presence of multiple ``null" hypotheses, all of which must be true simultaneously for two galaxy distributions to be the same. We apply the Bonferroni correction and adjust the confidence level accordingly, with $m=4$ as we test four properties in each host galaxy comparison. This means that, for a 95\% confidence level, a significant difference in galaxy properties now requires $p<0.05/4$. Thus no significant differences are seen between any other hosts and those of SLSNe~II; likely as a result of our relatively small sample.} We also note that this is despite the filter algorithm favouring blue, faint hosts for SLSN candidates (Sect. \ref{sec:data}).

Our full sample includes several events that spectroscopically resemble normal SNe~II more than they resemble SN~2008es. {It is possible that the 08es-like subgroup (i.e. those with a symmetric, broad H$\alpha$ line and weak absorption features) has host galaxies similar to SLSNe I, as suggested by \citet{inserra18}. There are, unfortunately, not enough such SNe} for a meaningful Anderson-Darling test. By eye the 08es-like SNe from ZTF occupy the high-mass portion of the SLSN I host distribution in Fig. \ref{fig:hosts} and seem to prefer higher SFRs than the SNe II or SLSNe IIn -- however, SN 2008es itself occurred in an extremely low-mass, low-SFR host. 

\section{Discussion}
\label{sec:disco}

\subsection{SLSNe II as a SN type} 

As stated above, we refer to our sample SNe as SLSNe~II, as opposed to events of similar luminosity with narrow Balmer lines, which we refer to as SLSNe~IIn. This follows \citet{inserra18} and is broadly consistent with the definition by \citet{galyam19}, who described the features of SLSNe~II in a similar way based on the four events known at the time. {SLSNe I can be distinguished from other events based on spectra alone \citep{quimby18}, and on occasion overlap with other H-poor SNe in luminosity \citep{gomez22}. Our sample of SLSNe II has, instead, been constructed simply by including all SNe in ZTF phase I with peak magnitudes $\leq-20$ mag, broad H$\alpha$ emission and a lack of strong narrow lines, therefore constituting a less robust and more heteronegeous group (see Sect. \ref{sec:spec}).} The sample includes SNe with peaks ranging from $-20$~mag to the extremely bright SN~2019zcr at $\sim-22.6$~mag. Few transients have reached a greater brightness, and these include SN~2015lh \citep{dong16}, likely a tidal disruption event \citep{leloudas16}, and three nuclear transients whose SN nature has not been ascertained \citep{kankare17}. 

We have, however, shown that with the exception of SN~2020jhm, objects in our sample spectroscopically resemble less luminous SNe~II. The relatively weak absorption lines, especially a lack of strong P~Cygni absorption in H$\alpha$, and the late emergence of the strong H$\alpha$ line in most cases, point toward SNe~II-L similar to SNe~1979C and 1998S \citep[e.g.][]{panagia80,fassia01}, although SN~2020yue is more similar to a Type II-P. The late-time spectra, however, lack strong forbidden lines of [O~{\sc i}] and [Ca~{\sc ii}] usually seen in SNe~II \citep[e.g.][]{dessart20}. A plateau or bump phase typical to normal SNe~II is not observed, and apart from two spectroscopically peculiar SLSNe~II, SNe~II-L decline faster at early times. {Instead, the light curves of SLSNe~II resemble those of other SLSNe.} This is in agreement with previous work on SLSNe~II \citep{inserra18}. Meanwhile, luminous SNe~II-L such as SN~2013fc \citep{kangas16} would be included in this sample if they were present. {Their absence} indicates that they are rare among ZTF targets. 

Observationally, we also note that only a minority of the sample resembles the prototypical SLSN~II, SN~2008es \citep{gezari09,miller09}, more than normal SNe~II; specifically, these 08es-like SNe exhibit more symmetric H$\alpha$ profiles (at all epochs studied here), typically brighter peaks and weaker absorption lines than the rest of the sample. Despite constituting three of the four previously studied SLSNe~II -- as SN~2013hx \citep{inserra18} can be considered a member of the other subgroup -- these SNe thus seem less common than the SLSNe~II more reminiscent of normal SNe~II. We point out, however, that these subgroups may be connected by a continuum of properties. {Since we define the sample based on a somewhat arbitrary luminosity limit, some degree of overlap or continuum with other H-rich SN subtypes may be expected as well.}

\subsection{Power sources}

Previous studies on SLSNe~II have suggested different mechanisms at work. \citet{miller09} favoured CSI with an opaque ejected shell as the dominant power source of SN~2008es, while \citet{inserra18}, with a sample of three SLSNe~II, suggested that a magnetar central engine is a good match to their light curves and temperatures. \citet{kornpob19} found the magnetar scenario to over-predict late-time fluxes of SN~2008es and unable to explain a NIR excess. They thus favoured the CSI model, but could not rule out the magnetar model with a declining fraction of trapped energy. We note that both \citet{inserra18} and \citet{kornpob19} used the bolometric light curve for their fits, while \texttt{MOSFiT} \citep{guillochon18} makes use of the colour information in the fitting as well. Even so, we find that the CSI and magnetar models included in \texttt{MOSFiT} are able to fit the light curves roughly equally well. $^{56}$Ni decay as the dominant power source can be ruled out, but from light curves alone it is difficult to distinguish between CSI and magnetar models. A similar conclusion was reached by \citet{chen22b} for a large sample of SLSNe~I from ZTF. Additionally, a few objects with UV data are not reproduced by \texttt{MOSFiT}.  

Some of the CSI results in Tables \ref{tab:resultsCSMNi2} and \ref{tab:resultsCSMNi0}, at face value, seem to require exotic or implausible scenarios. For example, in the $s=2$ case, SNe~2019aanx and 2019zcr have a CSM of tens of $\mathrm{M}_\odot$ and much more massive than the ejecta, requiring mass loss rates on the order of $0.1$ or $1\, \mathrm{M}_\odot$~yr$^{-1}$ depending on the wind velocity (i.e. an eruption, even if the density structure is $\propto \rho^{-2}$), and possibly fallback onto a nascent black hole resulting in a small ejecta mass. Meanwhile a $^{56}$Ni mass of $>10 \mathrm{M}_\odot$ is obtained for SN~2019uba with both $s$ values. While large ejecta masses themselves are plausible in e.g. pulsational pair instability SNe (PPISNe), $^{56}$Ni masses $\gtrsim4 \mathrm{M}_\odot$ likely require bona-fide PISNe \citep{kasen11,woosley17}.

There are, however, other observational indications in favour of CSI. As noted in Sect. \ref{sec:models}, it is possible that the model of \citet{chatz13} used in \texttt{MOSFiT} is too simplified to account for all the light curves, and a CSI power source should not be discarded based on the light curve modelling. For example, more detailed numerical models by \citet{sorokina16} produce very different light curves than the \citet{chatz13} model. It may be more fruitful to try to exclude the magnetar model: for SNe~2019cqc and 2019pud, we obtain high $^{56}$Ni masses of $\sim2.5 \mathrm{M}_\odot$ and $\sim0.9 \mathrm{M}_\odot$, respectively. We have, however, re-run both cases with $f_{\mathrm{Ni}} < 0.05$, and find fits of reasonable quality (by eye) without such high Ni masses. We have also re-run the CSI models for SN~2019uba, with a similar outcome. Thus the \texttt{MOSFiT} parameters should be treated with caution. We list the resulting parameters in Table \ref{tab:resultsRestrict}.

\begin{table*}
\centering    
\begin{minipage}{0.95\linewidth}
    \renewcommand{\arraystretch}{0.83}
    \centering
    \caption{Most important \texttt{MOSFiT} parameters and scores for models re-run with $f_{\mathrm{Ni}} < 0.05$. }
    \begin{tabular}{lcccccccc}
       Magnetar fit & log $B_\perp$ & $M_{\mathrm{NS}}$ & $P_{\mathrm{spin}}$ & log $f_{\mathrm{Ni}}$ & log $M_{\mathrm{ej}}$ & $\theta_{PB}$ & $v_{\mathrm{ej}}$ & Score\\
       & ($10^{14}$ G) & ($M_{\odot}$) & (ms) &  & ($M_{\odot}$) & (rad) & (km s$^{-1}$) & \\
        \hline
        SN~2019cqc & $0.4^{+0.3}_{-0.4}$ & $1.9^{+0.5}_{-0.6}$ & $6.6^{+1.1}_{-1.4}$ & $-1.9^{+0.5}_{-0.8}$ & $0.66\pm0.03$ & $0.5\pm0.2$ & $8020^{+30}_{-20}$ & 680\\
        SN~2019pud & $0.6\pm0.3$ & $1.8\pm0.5$ & $6.2\pm1.3$ & $-2.0^{+0.6}_{-0.7}$ & $0.51^{+0.04}_{-0.02}$ & $0.8\pm0.4$ & $15000^{+600}_{-1100}$ & 7.8\\
        \hline
        CSI fit & log $f_{\mathrm{Ni}}$ & log $M_{\mathrm{CSM}}$ & log $M_{\mathrm{ej}}$ & log $R_0$ & log $\rho$ & $v_{\mathrm{ej}}$ & $E_k$ & Score \\
       & & ($M_{\odot}$) & ($M_{\odot}$) & (AU) & (g cm$^{-3}$) & (km s$^{-1}$) & ($10^{51}$~erg) & \\
       \hline
        SN~2019uba (s=2) & $-2.6^{+0.7}_{-0.3}$ & $1.1^{+0.2}_{-0.4}$ & $-0.2^{+0.9}_{-0.5}$ & $2.5^{+0.3}_{-0.4}$ & $-13\pm1$ & $11000^{+700}_{-900}$ & $0.4^{+3.1}_{-0.3}$ & 68.9 \\
        SN~2019uba (s=0) & $-1.9\pm0.2$ & $0.83\pm0.06$ & $1.1^{+0.1}_{-0.2}$ & $2.2\pm0.1$ & $-12.4\pm0.2$ & $10000\pm400$ & $8\pm2$ & 72.0 \\
        \hline
    \end{tabular}
    \label{tab:resultsRestrict}
\end{minipage}
\end{table*}

The first indication of CSI comes from the UV photometry. As described in Sect. \ref{sec:lc_uv}, in four of the six SNe where an SED with UV data can be constructed, we observe an excess over a blackbody function in the UV. This was also seen in e.g. SN~1979C \citep{panagia80}, SN~2008es \citep{miller09} and SN~2010jl \citep{fransson14}. A forest of UV emission lines in SN~1979C was interpreted by \citet{fransson84} as being powered by excitation by X-rays from CSI. In spectra of SLSNe~I, which are mostly consistent with the magnetar model, the UV spectrum is instead heavily blanketed by absorption lines \citep[e.g.][]{yan17,yan18}. Additionally, absorption lines in SLSNe~II are relatively weak compared to SNe~II-P. \citet{branch00} attribute this to an additional contribution to the continuum emission from above the absorption layer, called "top-lighting". Such a scenario is better explained in the CSI model, as the central engine power source would necessarily be located below the absorption layer instead. Models by e.g. \citet{dessart22} also indicate both weakening absorption and increasing UV luminosity with increasing interaction power. The line profiles we observe, especially in the brighter, more 08es-like SNe, can be explained through CSI as well (see below). 

We note, however, that in the CSI scenario, the energy source is ultimately the kinetic energy of the ejecta. The neutrino-driven explosion mechanism may have problems with kinetic energies of more than a few~$\times10^{51}$~erg \citep{janka12}. According to the estimated UV-to-optical radiated energies (see Table \ref{tab:energy}), the explosion energies of the bright, 08es-like SNe must be $\gtrsim10^{51}$ (SN~2019aanx), $\gtrsim2\times10^{51}$ (SN~2019uba), or even $\gtrsim3\times10^{51}$ (SN~2019zcr) even with a 100 per cent conversion efficiency. Note that as these estimates were not extrapolated into unobserved epochs and wavelengths, the true radiated energy is larger still. 

The brightest SNe in the sample may thus need both CSI \emph{and} a central engine -- whether it be magnetar spin-down or possibly fallback accretion onto a black hole \citep{dexterkasen13}. A similar problem was noted by \citet{terreran17} for the extremely energetic OGLE-2014-SN-073. Even in a PPISN scenario, more than $5\times10^{51}$~erg of kinetic energy becomes a problem without a magnetar \citep{woosley17}. More detailed studies will be necessary to constrain the contribution and nature of this additional power source. This can be done through e.g. late-time radio follow-up and possible detections of young pulsar wind nebulae \citep{omand18,law19,eftekhari21}. The kinetic energies from the CSI models, estimated as $E_k = 0.3M_{ej}v_{ej}^2$, also often reach $>3\times10^{51}$~erg (see Tables \ref{tab:resultsCSMNi2} and \ref{tab:resultsCSMNi0}), tentatively suggesting an additional power source as well; but the uncertainties are often large enough to allow $<2\times10^{51}$~erg. Only three SNe (not including the SN with the highest radiated energy, SN~2019zcr!) require $E_k >3\times10^{51}$~erg in \emph{both} the $s=0$ and $s=2$ models when uncertainties are taken into account. 

Finally, we point out that SN~2020hgr, the slowest-evolving SN in our sample (see Sect. \ref{sec:lc}), is superficially similar to some PISN models. \citet{kasen11} showed that spectra of PISNe with hydrogen-rich $150$--$250~ \mathrm{M}_\odot$ progenitors can also look similar to those of normal SNe~II and SN~2020hgr. However, the light curves of such models quickly rise to a plateau phase and/or have a main peak at hundreds of days post-explosion. PISNe from $80$--$100~ \mathrm{M}_\odot$ helium stars do photometrically resemble that of SN~2020hgr, but the spectrum of SN~2020hgr is hydrogen-dominated until late times, while CCSNe with very small hydrogen masses (type IIb) eventually develop strong helium features. These models tentatively argue against a PISN scenario, but do not rule it out entirely.

\subsection{Line profiles and CSM structure} 

\citet{taddia20} modeled the line profile in SN~2013L, a SN~IIn which exhibited both narrow/intermediate and broad H$\alpha$ components. They showed that the shape of the broad component can be reproduced with a spherically symmetric CSM shell. A cool, dense shell (CDS) forms between the forward and reverse shocks \citep{cf94,cf17}. Broad emission lines would originate behind the radiative forward shock and would, without electron scattering, result in a boxy profile. However, \citet{taddia20} showed that a high optical depth for electron scattering ($\tau_e$) in the unshocked, ionized CSM, combined with occultation of the receding side, would produce an emission profile with red-wing suppression but no broad P~Cygni absorption, similar to what is seen in most SLSNe~II. As $\tau_e$ increases further, this profile would become symmetric, as seen in the brightest SLSNe II. 

The narrow/intermediate line profile in such a scenario originates in the ionized, unshocked CSM that is also responsible for the electron scattering. If $\tau_e$ is high enough ($\gtrsim30$), the existence of a non-dominant narrow electron scattering component such as in SN~2013L, even if easily visible in emission at a lower $\tau_e$, can be hidden \citep{taddia20}. In the case of an asymmetric line profile and lower $\tau_e$, the narrow emission component must be intrinsically weak. A low density in the unshocked region would result in a weak narrow component -- requiring an extended ionized CSM. A weak narrow feature, or a narrow P Cygni profile from optically thin outer CSM, may escape detection without high-resolution spectra, as seen e.g. in SNe~2010jl and the very similar 2015da at late times \citep{zhang12,fransson14,tartaglia20}. Stronger interaction would result in both a more luminous SN and simultaneously a higher $\tau_e$, as any extended CSM would be ionized further out. 

Late-time multi-component H$\alpha$ profiles were seen in PS15br and SN~2013hx \citep{inserra18}; they can in principle result from the shell interaction described above \citep[][]{taddia20}. However, one component in this case is the smeared-out narrow profile, whereas in PS15br a narrow P~Cygni feature was not seen even in high-resolution spectra, making this scenario unlikely, and SN~2013hx showed a three-component profile. They thus likely still require asymmetric CSM, but for our SNe, which lack such line profiles at late times, this is not necessary. The lack of strong forbidden metal lines in late-time spectra can simply be due to not being truly nebular at $\lesssim+300$~d. The density at the emitting region is still high enough that atoms/ions are collisionally de-excited instead of emitting in forbidden lines. The red wing is often suppressed, implying occultation of the far side -- this also argues that the spectra are not nebular. Longer follow-up campaigns are needed to study SLSNe~II in the nebular phase.

The scenario described above may not be required for all SLSNe~II. Some show P Cygni profiles in H$\alpha$ similar to less luminous SNe~II (or, for SN~2019pud, no clear absorption trough but a suppressed blue wing, presumably also from absorption in H-rich ejecta), which indicate a line of sight into the ejecta and are not expected to be seen through an optically thick CDS. In SN~2020yue, this is seen very early. A clumpy CDS can result in optically thin gaps in the CDS \citep{smith08}, but it is also possible that the CSM is overrun by the ejecta in an early stage, e.g. if the CSM is disk-shaped \citep{mcdowell18}. 

\citet{moriya12} proposed a model in which the forward shock breaks out of an optically thick shell or wind CSM, resulting in a broadened light curve peak. At early times, this results in a blue, featureless spectrum, followed by the broad emission lines post-peak. If the outer layers of the CSM are optically less thick, they would be unshocked but ionized and result in a SN~IIn, while a SN~II-L could result if the CSM density is roughly constant: very little unshocked CSM remains after this breakout and photons originate from the shocked CSM and the ejecta. We do not see early bumps in our light curves (see Figs. \ref{fig:george2} and \ref{fig:george1}) similar to what e.g. \citet{angus19} observe in some SLSNe~I, possibly associated with shock cooling and an extended progenitor. In most cases we cannot exclude them either, but their absence is consistent with the main peak being associated with a breakout from the CSM shell. A similar scenario has been proposed for normal SNe~II \citep{forster18}.

A combination of effects may be at play. The brightest SLSNe~II with broad symmetric emission, and possibly others, can be explained through electron scattering in the unshocked CSM. A range of $\tau_e$ values can produce different H$\alpha$ lines, symmetric profiles with a very high $\tau_e$ and profiles similar to SN~2019gsp or SN~2020bfe possibly with a somewhat lower $\tau_e$. Other SLSNe~II, such as SN~2020hgr, which seems to require CSI based on its UV excess but whose spectrum is extremely similar to SNe~II-L, may require dense CSM confined to small radii, possibly in a disk shape. Normal wind mass loss \citep[$\lesssim 10^{-6}~\mathrm{M}_\odot$~yr$^{-1}$ in RSGs according to][]{beasor20} would not significantly contribute to the emission.

CSI has been argued to be required in the early epochs of many if not all SNe~II \citep[][]{morozova18}; its signatures can be seen in the radio light curves of SNe II-L \citep{lf88} and in the UV spectrum of SN~1979C \citep{fransson84}. An increasing amount of CSM can result in light curves with shorter plateaus and brighter, broader peaks \citep{moriya11}, even including absolute magnitudes of $\sim-22$~mag for pre-SN mass loss rates $\gtrsim0.1~\mathrm{M}_\odot$~yr$^{-1}$. We thus suggest a continuum of recent mass loss from SNe~II-P through SNe~II-L to some SLSNe~II. At least the SNe spectroscopically similar to SN~2008es seem to additionally require a more extended CSM and thus may be a separate group. Detailed numerical modelling of SLSNe~II to determine the properties of the CSM is outside the scope of this study. 

\subsection{Clues on SLSN~II progenitors}

As stated in Sect. \ref{sec:data}, our sample criteria are matched by 14 of the 69 hydrogen-rich SLSNe followed up in the ZTF phase I. This would imply a fraction of $0.20^{+0.05}_{-0.06}$ \citep{gehrels86} out of all hydrogen-rich SNe with $M_g < -20$~mag have broad emission lines without narrow ones. However, as we point out in Sect. \ref{sec:lccomp}, there is some overlap in photometric properties between SLSNe~II and IIn in ZTF. It is possible, as stated in Sect. \ref{sec:data}, that since some SLSNe~IIn in ZTF only have pre-peak spectra, broad lines might have appeared later and replaced the earlier narrow lines (and the latter may in rare cases be from the host). Therefore the fraction should be considered a lower limit: SNe similar to our sample make up $>14$ per cent of the transients classified as SLSNe~II by ZTF. If spectroscopically 08es-like SNe, i.e. those with the highest $\tau_e$ and strongest interaction, are considered separately, we can similarly imply a fraction of 3/69, i.e. $0.04\pm0.03$. This should again be considered a lower limit, i.e. $>1$~per cent. This is in line with the rarity of such SNe in the literature even compared to other SLSNe \citep{inserra18}.

The properties of the progenitor systems of these SNe -- i.e. mass loss history and initial mass -- must be unusual, even for SLSNe. The mass loss history can be affected by a binary companion, the metallicity of the progenitor and/or its initial mass \citep[e.g.][]{smith14}. Although light curves alone cannot rule out magnetars, our results indicate CSI is required by most if not all SLSNe~II; SLSNe~IIn, on the other hand, are clearly primarily powered by CSI. If CSI is responsible for both, the density profile of the CSM and thus mass loss history must be different -- but weaker narrow emission lines or narrow P Cygni profiles can escape detection without high-resolution spectra. The host galaxies of SLSNe~II are quite similar to those of SLSNe~IIn (see Sect. \ref{sec:hosts}), which suggests that their environments and metallicities are similar as well (but our sample size remains rather small). In such a case a difference in progenitor mass and/or a binary companion could be causing the different mass loss histories of SLSNe~II and IIn. In the \citet{moriya12} and \citet{taddia20} models, an asymmetric CSM is not required. Instead of (or in addition to) mass loss through binary interaction, the ejection of a spherically symmetric shell in an eruption close to the death of the progenitor star may result in the dense CSM that would produce the observed line profiles. 

Eruptive mass loss seems to be necessary even in many normal SNe~II, where CSM masses may reach $\gtrsim0.5\, \mathrm{M}_\odot$ \citep{morozova18} and CSM radii have been argued to be on the order of $1000\, R_\odot$, i.e. not much larger than the progenitors themselves. \citet{moriya11} also suggested strong mass loss in RSGs and yellow supergiants (YSGs) just before their deaths. In SLSNe~IIn, on the other hand, a longer-lasting, strong wind or a series of eruptive events \citep[such as pulsational pair instability;][]{woosley07} may be responsible for a more extended dense CSM. A possible mechanism for eruptions in the very late stages of RSG evolution is the so-called wave-driven mass loss \citep{qs12,sq14}, which can unbind up to $\sim10\, \mathrm{M}_\odot$ in the last months or years before explosion after carbon burning. 

The progenitors of normal SNe II-P are established as RSGs of roughly 8--$17\, \mathrm{M}_\odot$ \citep{smartt09}, while the progenitors of SNe~II-L are less well known, but consistent with initial masses of 15--$20\, \mathrm{M}_\odot$ \citep{vandyk99,faran14,kangas16,kangas17a}. Very massive luminous blue variable stars, on the other hand, are connected to SNe~IIn \citep[e.g.][]{smith10,smith11,mauerhan13,taddia13,fransson14}. If SNe~II-P, II-L and some SLSNe~II are connected by a continuum of increasing mass lost through a similar eruptive mechanism, the rare progenitors of the latter may be the most massive RSGs or YSGs. The diversity within the sample may indicate multiple progenitor scenarios, though. Other SLSNe~II, including those similar to SN~2008es, may require a more extended, low-density CSM as the location of electron scattering. 
Based on our host galaxy modelling, the 08es-like SNe -- which may require central engines as well as CSI -- may also favour lower metallicities than the rest, which is the case for SLSNe~I where magnetar engines are the most popular scenario \citep{perley16,Schulze2018a}.


\section{Summary and conclusions}
\label{sec:concl}


We have examined the light curves and spectra of a sample of 14 SLSNe~II from ZTF phase I that exhibit broad Balmer line emission without strong narrow lines typical to (SL)SNe~IIn. This is the largest such sample to date. We have used light-curve models to attempt to constrain the power sources responsible for the luminosity of SLSNe~II. Based on this work, we draw the following conclusions.
\begin{itemize}
    \item The spectra of several SLSNe~II are very similar to those of SNe~II-L. Broad, asymmetric Balmer line emission is accompanied by weak or non-existent P~Cygni absorption and metal lines typical to SNe~II. Photometrically, these SLSNe evolve slower than normal SNe~II-L and do not clearly exhibit the typical plateau/bump phase followed by a drop to $^{56}$Co decay tail, instead resembling SLSNe~I.
    \item Other SLSNe~II include three very luminous ($M_g \leq-21.7$~mag, even $-22.6$~mag, at peak and requiring at least $10^{51}$~erg in radiated energy) SNe that resemble the prototypical event SN~2008es more than normal SNe~II. H$\alpha$ emission is symmetric until late times and absorption lines are weak. A close resemblance to SN~2008es is, however, far from ubiquitous among SLSNe~II. The sample also includes two fast-declining SNe that exhibit spectroscopic features not present in the two aforementioned groups.
    \item Light-curve models invoking a magnetar engine \citep{kb10} or CSI \citep{chatz13} are roughly equally successful in reproducing the observed evolution, and some SLSNe~II with UV observations are difficult for both. Only a pure $^{56}$Ni-powered model can be excluded.
    \item However, we observe an excess in the UV compared to a blackbody in several cases including a wide range of photometric and spectral properties, which is likely due to a forest of emission lines from various species ionized by X-rays from CSI. At least these SNe seem to be CSM-powered. It is also possible, however, that the extreme radiated energies of the brightest SLSNe~II, $\gtrsim2\times10^{51}$ or even $\gtrsim3\times10^{51}$~erg, require a central engine as well as CSI.
    \item The emission lines of the brightest, 08es-like SLSNe~II can be explained through interaction with a dense CSM, observed through a screen of ionized, unshocked CSM optically thick to electron scattering. SLSNe~II spectroscopically more similar to normal SNe~II may involve a dense CSM confined to small radii of the progenitor star. Eruptive mass loss has been argued to be important in SNe~II, especially II-L; thus some SLSNe~II may be connected to normal SNe~II through a continuum of pre-SN mass loss. 
    \item SLSNe~II without narrow emission lines comprise roughly 20 per cent of all hydrogen-rich SLSNe followed up by ZTF. This rarity even compared to other SLSNe indicates highly unusual progenitor stars and/or mass loss histories. The host galaxies of SLSNe~II strongly overlap with those of SLSNe~IIn, indicating a similar environment; differences in mass loss history may thus be connected to initial mass rather than metallicity.
    \item All observed late-time spectra of SLSNe~II, including SN~2013hx and PS15br \citep{inserra18}, lack strong forbidden metal lines typical to normal SNe~II at similar phases; these SNe are likely not yet nebular at $\sim+300$~d. No multi-component line profiles similar to SN~2013hx and PS15br are seen in our sample, indicating that interaction with \textit{asymmetric} CSM is not required by most late-time spectra.
\end{itemize}

\section*{Acknowledgements}
We thank the anonymous referee for suggestions that helped to improve the paper.

This study is based on observations obtained with the Samuel Oschin Telescope 48-inch and the 60-inch Telescope at the Palomar Observatory as part of the Zwicky Transient Facility project. ZTF is supported by the National Science Foundation under Grants No. AST-1440341 and AST-2034437 and a collaboration including current partners Caltech, IPAC, the Weizmann Institute for Science, the Oskar Klein Center at Stockholm University, the University of Maryland, Deutsches Elektronen-Synchrotron and Humboldt University, the TANGO Consortium of Taiwan, the University of Wisconsin at Milwaukee, Trinity College Dublin, Lawrence Livermore National Laboratories, IN2P3, University of Warwick, Ruhr University Bochum, Northwestern University and former partners the University of Washington, Los Alamos National Laboratories, and Lawrence Berkeley National Laboratories. Operations are conducted by COO, IPAC, and UW. SED Machine is based upon work supported by the National Science Foundation under Grant No. 1106171. The ZTF forced-photometry service was funded under the Heising-Simons Foundation grant \#12540303 (PI: Graham). This work uses the GROWTH Followup Marshal \citep{kasliwal19} and was supported by the GROWTH project funded by the National Science Foundation under Grant No 1545949.

This study is based partially on observations made with the Nordic Optical Telescope, owned in collaboration by the University of Turku and Aarhus University, and operated jointly by Aarhus University, the University of Turku and the University of Oslo, representing Denmark, Finland and Norway, the University of Iceland and Stockholm University at the Observatorio del Roque de los Muchachos, La Palma, Spain, of the Instituto de Astrof\'{i}sica de Canarias. The data presented here were obtained in part with ALFOSC, which is provided by the Instituto de Astrof\'{i}sica de Andalucia (IAA) under a joint agreement with the University of Copenhagen and NOT. The Liverpool Telescope is operated on the island of La Palma by Liverpool John Moores University in the Spanish Observatorio del Roque de los Muchachos of the Instituto de Astrof\'{i}sica de Canarias with financial support from the UK Science and Technology Facilities Council. The William Herschel Telescope is operated on the island of La Palma by the Isaac Newton Group of Telescopes in the Spanish Observatorio del Roque de los Muchachos of the Instituto de Astrof\'{i}sica de Canarias. 

Some of the data presented herein were obtained at the W. M. Keck Observatory, which is operated as a scientific partnership among the California Institute of Technology, the University of California and the National Aeronautics and Space Administration. The Observatory was made possible by the generous financial support of the W. M. Keck Foundation. The authors wish to recognize and acknowledge the very significant cultural role and reverence that the summit of Maunakea has always had within the indigenous Hawaiian community. We are most fortunate to have the opportunity to conduct observations from this mountain. 

T.K. acknowledges support from the Swedish National Space Agency and the Swedish Research Council. S. Schulze acknowledges support from the G.R.E.A.T research environment, funded by {\em Vetenskapsr\aa det}, the Swedish Research Council, project number 2016-06012. R. L. acknowledges support from a Marie Sk\l odowska-Curie Individual Fellowship within the Horizon 2020 European Union (EU) Framework Programme for Research and Innovation (H2020- MSCA-IF-2017-794467). T.-W. C. acknowledges the EU Funding under Marie Sk\l{}odowska-Curie grant H2020-MSCA-IF-2018-842471. L.T. acknowledges support from MIUR (PRIN 2017 grant 20179ZF5KS). A.G-Y.’s research is supported by the EU via ERC grant No. 725161, the ISF GW excellence center, an IMOS space infrastructure grant and BSF/Transformative and GIF grants, as well as the Andr\'{e} Deloro Institute for Advanced Research in Space and Optics, the Schwartz/Reisman Collaborative Science Program and the Norman E Alexander Family M Foundation ULTRASAT Data Center Fund, The Kimmel center for Planetary Sciences, Minerva and Yeda-Sela; A.G-Y. is the incumbent of the The Arlyn Imberman Professorial Chair.

\section*{Data availability}

The photometry data used in this study are included as supplementary material; raw images used in ZTF photometry are publicly available through ZTF Data Release 10 at \url{https://www.ztf.caltech.edu/ztf-public-releases.html}. All output corner plot files and figures from light-curve modelling are also included as supplementary material. Reduced spectra used in this study will be publicly available through WISEREP (\url{https://www.wiserep.org/}). The derived data generated in this research will be shared on reasonable request to the corresponding author.

Optical/UV \emph{Swift} photometry data used in this paper are available publicly on the NASA Swift Data Archive (\url{https://heasarc.gsfc.nasa.gov/}). X-ray count rates are publicly available on \emph{Swift} X-ray Telescope (XRT) website (\url{https://www.swift.ac.uk/user_objects/}). Similarly, the host galaxy data used in this paper were obtained from public data repositories.




\bibliographystyle{mnras}
\bibliography{kangas_arXiv_20220722} 



\appendix

\section{Logs of observations}
\label{app:logs}

\begin{table*}
\centering    
\begin{minipage}{0.95\linewidth}
    \centering    
    \caption{The first 10 lines of the log of spectroscopic observations used in this paper. The full table is available in supplementary material. }
    \begin{tabular}{lcccccccc}
       SN & Julian date & Phase & Telescope & Instrument & Exposure time & Slit & Grism/grating \\
       & (d) & (d) & & & (s) & \arcsec & \\
        \hline
        SN~2018jkq & 2458462.700 & -12 & P60 & SEDM & 2250 & IFU & -\\
        SN~2018jkq & 2458471.737 & -4 & P60 & SEDM & 1200 & IFU & -\\ 
        SN~2018jkq & 2458472.776 & -3 & P60 & SEDM & 1200 & IFU & -\\ 
        SN~2018jkq & 2458473.652 & -2 & P60 & SEDM & 1200 & IFU & -\\ 
        SN~2018jkq & 2458479.615 & 3 & P200 & DBSP & 600 & 2.0 & 600/4000+316/7500\\ 
        SN~2018jkq & 2458480.763 & 4 & P60 & SEDM & 1200 & IFU & - \\ 
        SN~2018jkq & 2458484.341 & 7 & WHT & ISIS & 800 & 1.5 & R300B+R316R\\ 
        SN~2018jkq & 2458490.702 & 13 & P200 & DBSP & 900 & IFU & 600/4000+316/7500\\ 
        SN~2018jkq & 2458667.051 & 170 & Keck & LRIS & 900 & IFU & 400/3400+400/8500\\
        SN~2018lqi & 2458762.858 & -18 & P200 & DBSP & 900 & 1.5 & 600/4000+316/7500\\ 
        \hline
    \end{tabular}
    \label{apptab:Log}
\end{minipage}
\end{table*} 

\begin{table*}
\centering    
\begin{minipage}{0.95\linewidth}
    \centering    
    \caption{The first 20 lines of the table of photometry used in this paper. The full table is available in supplementary material. An apparent magnitude of 99 denotes a non-detection.}
    \begin{tabular}{lcccccc}
       SN & Julian date & Filter & App. magnitude & Error & Upper limit & Instrument \\
       & (d) & & (mag) & (mag) & (mag) & \\
        \hline
        SN2018jkq & 2458426.7547 & $r$ & 99 & 99 & 20.41 & P48+ZTF \\
        SN2018jkq & 2458426.8254 & $g$ & 99 & 99 & 20.08 & P48+ZTF \\
        SN2018jkq & 2458427.7796 & $i$ & 99 & 99 & 19.13 & P48+ZTF \\
        SN2018jkq & 2458427.7829 & $i$ & 99 & 99 & 18.85 & P48+ZTF \\
        SN2018jkq & 2458429.7746 & $r$ & 99 & 99 & 20.71 & P48+ZTF \\
        SN2018jkq & 2458431.7597 & $i$ & 99 & 99 & 19.05 & P48+ZTF \\
        SN2018jkq & 2458431.7630 & $i$ & 99 & 99 & 19.16 & P48+ZTF \\
        SN2018jkq & 2458432.7688 & $r$ & 99 & 99 & 20.20 & P48+ZTF \\
        SN2018jkq & 2458435.7287 & $g$ & 99 & 99 & 20.25 & P48+ZTF \\
        SN2018jkq & 2458435.7327 & $i$ & 99 & 99 & 19.26 & P48+ZTF \\
        SN2018jkq & 2458435.7685 & $r$ & 99 & 99 & 19.80 & P48+ZTF \\
        SN2018jkq & 2458438.8190 & $r$ & 99 & 99 & 17.46 & P48+ZTF \\
        SN2018jkq & 2458438.8583 & $g$ & 99 & 99 & 16.97 & P48+ZTF \\
        SN2018jkq & 2458441.7384 & $g$ & 99 & 99 & 19.53 & P48+ZTF \\
        SN2018jkq & 2458441.7752 & $r$ & 99 & 99 & 19.41 & P48+ZTF \\
        SN2018jkq & 2458447.8300 & $r$ & 99 & 99 & 19.46 & P48+ZTF \\
        SN2018jkq & 2458450.7496 & $r$ & 99 & 99 & 18.14 & P48+ZTF \\
        SN2018jkq & 2458450.7550 & $g$ & 99 & 99 & 18.53 & P48+ZTF \\
        SN2018jkq & 2458450.8014 & $r$ & 99 & 99 & 15.62 & P48+ZTF \\
        SN2018jkq & 2458456.6831 & $r$ & 18.94 & 0.08 & 20.25 & P48+ZTF \\
        \hline
    \end{tabular}
    \label{apptab:Phot}
\end{minipage}
\end{table*} 

Full logs of all spectroscopic observations and optical and UV photometry (including upper limits) used in this study are available as supplementary material. We present samples of these logs in Tables \ref{apptab:Log} and \ref{apptab:Phot}, respectively.

\section{X-ray limits}
\label{app:Xray}

We have obtained the count-rate limits for our targets on the \emph{Swift} X-ray Telescope (XRT) website\footnote{\url{https://www.swift.ac.uk/user_objects/}}. A total of nine SNe out of the sample have associated XRT observations after the discovery of the SN; these are listed in Table \ref{apptab:XRT}. No SLSN source was detected in these observations. We therefore list the upper limits in the table in terms of count rate and flux over the 0.2--10 keV band of XRT. Count rates were converted to fluxes using \texttt{WebPIMMS}\footnote{\url{https://heasarc.gsfc.nasa.gov/cgi-bin/Tools/w3nh/w3nh.pl}}, assuming host hydrogen column density $n_{H,\mathrm{host}} = 2\times10^{21}$~erg (the least constraining case in our light-curve models; see Sect. \ref{sec:modsetup}) and an Astrophysical Plasma Emission Code (APEC) model with a temperature of 19 keV, similar to the X-ray spectrum of SN~2010jl \citep{chandra15}. Galactic hydrogen column densities were obtained from the NASA HEASARC \texttt{nH} tool\footnote{\url{https://heasarc.gsfc.nasa.gov/cgi-bin/Tools/w3nh/w3nh.pl}}.

The $3\sigma$ upper limits in the 0.2--10 keV range from the XRT archive are typically on the order of a few $\times 10^{-13}$~erg~s$^{-1}$~cm$^{-2}$ or even higher, an order of magnitude less constraining than that determined for SN~2008es \citep{gezari09}. Depending on the target redshift, these translate into limits on the 0.2--10 keV luminosity between $\lesssim10^{42}$ and $\lesssim10^{44}$~erg~s$^{-1}$ (SN~2020jhm at 480~d and SN~2019zcr at $<50$~d, respectively). These limits are orders of magnitude higher than the observed luminosity of the strongly interacting, relatively nearby and almost superluminous SN~2010jl between 50 and 1300~d \citep{chandra15}, from $6\times10^{39}$ to $1.3\times10^{40}$~erg~s$^{-1}$. Thus the X-ray non-detections do not exclude CSI as the dominant power source.

\begin{table}
    \centering
    \caption{Upper limits ($3\sigma$) of count rate and unabsorbed flux in the 0.2--10 keV band of XRT for sample SNe with associated X-ray observations. Epochs are relative to the rest-frame $g$-band peak.}
    \begin{tabular}{lccc}
       SN & Epoch & Count rate & Flux\\
       & (d) & (10$^{-3}$ ct s$^{-1}$) & (erg s$^{-1}$ cm$^{-2}$) \\
        \hline
        SN~2019kwr & 733.5 & $<7.8$ & $<4.6 \times 10^{-13}$ \\ 
        SN~2019cqc & 26.9 & $<3.4$ & $<2.2 \times 10^{-13}$ \\
          & 40.5 & $<3.1$ & $<2.0 \times 10^{-13}$\\
          & 47.0 & $<3.2$ & $<2.1 \times 10^{-13}$ \\
          & 53.0 & $<2.6$ & $<1.7 \times 10^{-13}$\\
          & 815.1 & $<2.1$ & $<1.4 \times 10^{-13}$\\
          & 819.2 & $<6.2$ & $<4.0 \times 10^{-13}$ \\
        SN~2019xfs & 33.4 & $<4.2$ & $<2.8 \times 10^{-13}$ \\
          & 39.1 & $<3.5$ & $<2.3 \times 10^{-13}$ \\
          & 45.9 & $<6.7$ & $<4.4 \times 10^{-13}$ \\
          & 60.8 & $<6.7$ & $<4.4 \times 10^{-13}$ \\
          & 67.3 & $<3.9$ & $<2.6 \times 10^{-13}$ \\
          & 73.2 & $<3.5$ & $<2.3 \times 10^{-13}$\\
          & 621.8 & $<3.4$ & $<2.2 \times 10^{-13}$\\
        SN~2019pud & 14.4 & $<8.8$ & $<5.6 \times 10^{-13}$\\
          & 18.0 & $<5.9$ & $<3.7 \times 10^{-13}$ \\
        SN~2019uba & 9.5 & $<9.8$ & $<5.6 \times 10^{-13}$\\ 
          & 13.2 & $<4.4$ & $<2.5 \times 10^{-13}$\\
          & 15.7 & $<3.6$ & $<2.1 \times 10^{-13}$\\
        SN~2019zcr & $-13.0$ & $<17.6$ & $<1.0 \times 10^{-12}$ \\ 
          & $-5.5$ & $<6.1$ & $<3.5 \times 10^{-13}$\\
          & 11.0 & $<8.6$ & $<5.0 \times 10^{-13}$\\
          & 15.6 & $<9.0$ & $<5.2 \times 10^{-13}$\\
          & 19.4 & $<5.2$ & $<3.0 \times 10^{-13}$\\
          & 23.0 & $<6.9$ & $<4.0 \times 10^{-13}$\\
          & 28.6 & $<9.9$ & $<5.7 \times 10^{-13}$\\
          & 30.4 & $<19.9$ & $<1.1 \times 10^{-12}$\\
          & 34.4 & $<32.3$ & $<1.9 \times 10^{-12}$\\
          & 36.9 & $<20.9$ & $<1.2 \times 10^{-12}$ \\
          & 38.7 & $<13.9$ & $<8.0 \times 10^{-13}$\\
          & 42.9 & $<12.5$ & $<7.2 \times 10^{-13}$\\
          & 48.1 & $<24.4$ & $<1.4 \times 10^{-12}$\\
        SN~2020hgr & $-5.3$ & $<6.8$ & $<4.1 \times 10^{-13}$\\
          & 0.9 & $<9.5$ & $<5.7 \times 10^{-13}$\\
          & 10.7 & $<18.2$ & $<1.1 \times 10^{-12}$\\
          & 22.9 & $<34.5$ & $<2.1 \times 10^{-12}$\\
          & 28.0 & $<7.8$ & $<4.7 \times 10^{-13}$\\
          & 39.2 & $<5.1$ & $<3.1 \times 10^{-13}$\\
        SN~2020jhm & 24.3 & $<8.5$ & $<5.4 \times 10^{-13}$\\
          & 28.1 & $<8.8$ & $<5.6 \times 10^{-13}$\\
          & 37.8 & $<8.2$ & $<5.2 \times 10^{-13}$\\
          & 480.2 & $<2.8$ & $<1.7 \times 10^{-13}$\\
        SN~2020yue & 2.7 & $<6.3$ & $<3.7 \times 10^{-13}$\\
          & 8.5 & $<8.5$ & $<5.0 \times 10^{-13}$ \\
          & 14.2 & $<13.0$ & $<7.6 \times 10^{-13}$\\
          & 21.7 & $<6.8$ & $<4.0 \times 10^{-13}$\\
          & 25.6 & $<6.2$ & $<3.6 \times 10^{-13}$\\
          & 266.5 & $<4.2$ & $<2.4 \times 10^{-13}$\\
        \hline
    \end{tabular}
    \label{apptab:XRT}
\end{table} 

\section{Notes on MOSFiT magnetar model}
\label{app:model}

\texttt{MOSFiT} includes an alternative model for magnetar central engines, labeled {\tt slsn}. This model includes various constraints \citep[for details, see][]{nicholl17} and a modified SED below 3000~\AA, emulating the effect of line blanketing common in the UV spectra of SLSNe I \citep[e.g.][]{yan17}. In the \texttt{magni} model, the SED is simply a blackbody. As established in Sect. \ref{sec:lc_uv}, neither a simple blackbody nor this modified function with line blanketing can fit four of the six SEDs in our sample. No major deficit from line blanketing is seen. Therefore we have used the \texttt{magni} model in our fits.

For setting a minimum ejecta mass, we conservatively assumed the initial mass of the progenitor was $\gtrsim14~\mathrm{M}_\odot$; the lowest progenitor mass ascribed to an observed magnetar is, to our knowledge, $\sim17~\mathrm{M}_\odot$ \citep{davies09}. Typically, the initial masses ascribed to millisecond magnetar progenitors are $\gtrsim30~\mathrm{M}_\odot$ \citep[e.g.][]{gaensler05,heger05,ok14}. As some hydrogen must be left at the end of the star's life \citep[for SNe II-L, $\sim1~\mathrm{M}_\odot$ according to][]{bb93}, we use the \texttt{PARSEC} stellar evolution models\footnote{http://stev.oapd.inaf.it/cgi-bin/cmd} \citep{bressan12} to find the final He/CO core mass of an initially $14~\mathrm{M}_\odot$ star assuming Solar metallicity\footnote{The core mass increases slightly with lower metallicity; thus assuming Solar metallicity here is the least constraining choice.}: $\sim4.5~\mathrm{M}_\odot$. Assuming the entire core mass is still present along with the hydrogen envelope, this results in a final progenitor mass of $\gtrsim5.5~\mathrm{M}_\odot$. Assuming the maximum mass of the magnetar born in the collapse is $\sim2.5~\mathrm{M}_\odot$ \citep[][]{shibata19}, the minimum ejecta mass was therefore set at $3~\mathrm{M}_\odot$. 


\bsp	
\label{lastpage}
\end{document}